\documentclass{WileyMSP-template}
\usepackage{amsmath}
\usepackage[dvipsnames]{xcolor}
\begin{document}

\pagestyle{fancy}
\rhead{\includegraphics[width=2.5cm]{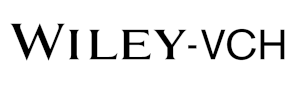}}

\title{Nonlinear Distortion Equalization in Multi-Span Optical Links Via a Feed-Forward Photonic Neural Network}

\maketitle

\author{Emiliano Staffoli}
\author{Elisabetta Ferri}
\author{Stefano Gretter}
\author{Lorenzo Pavesi*}

\dedication{}

\begin{affiliations}
Emiliano Staffoli, Elisabetta Ferri, Stefano Gretter, Prof. Lorenzo Pavesi\\
via sommarive, 14 - Povo (TN), 38123, Trento, Italy\\
Email Address: lorenzo.pavesi@unitn.it
\end{affiliations}



\begin{abstract}

Linear and nonlinear distortions in optical communication signals are equalized using an integrated feed-forward Photonic Neural Network (PNN). The PNN is based on a linear stage made of an 8-tap Finite Impulse Response (FIR) filter, featuring tunable amplitude and phase weights at each tap, and of a nonlinear stage achieved through the square modulus operation at the end-of-line photodetector. Within an Intensity Modulation/Direct Detection (IMDD) system, the PNN is applied to 2-level Pulse Amplitude Modulated (PAM2) optical signals undergoing multi-span propagation. Each 50 km segment includes fiber transmission, optical power restoration, and optional chromatic dispersion compensation via a Tunable Dispersion Compensator. Positioned at the receiver, the PNN enables fully optical signal processing with minimal latency and power consumption. Experimental validation is conducted using a Silicon-On-Insulator device operating on 10 Gbps signals. It demonstrates chromatic dispersion equalization over distances up to 200 km and self-phase modulation (with dispersion removed) up to 450 km. Simulations explore PNN adaptation for 100 Gbps modulations and its potential for cross-phase modulation equalization.

\end{abstract}


\section{Introduction}
Digital Signal Processing (DSP) has brought about a paradigm shift in optical telecommunications, enabling advanced signal manipulation techniques that were previously unattainable \cite{zhong2018digital, guifang2009recent, liu2014digital, zhou2014advanced}. These systems rely on processors built using Application-Specific Integrated Circuits (ASICs), which are tasked with a wide array of functions, such as clock and data recovery \cite{sonntag2006digital} and error correction \cite{chang2010forward}. Central to DSP functionality are sophisticated algorithms designed to address channel equalization \cite{huang2022performance}, compensating for signal impairments arising from both linear distortions (e.g., Chromatic Dispersion, CD) and nonlinear effects (e.g., Self- and Cross-phase modulation, SPM, XPM) \cite{agrawal2013nonlinearCh1}. 

The telecommunication and data communication markets are growing at an unprecedented pace with an increased demand for higher bandwidth and faster data rates. These requests pose an increasing strain on DSP capabilities \cite{zhong2018digital}. Expanding bandwidth requirements necessitate more advanced processing power, significantly raising energy consumption and operational costs \cite{frey2017estimation}. This challenge is particularly pronounced in coherent optical transceivers, which are designed to handle high data rates, long distances and system complexities \cite{cheng2019comparison}. On the other hand, Intensity Modulation/Direct Detection (IMDD) transceivers offer a more straightforward and cost-effective alternative, trading off bandwidth efficiency for affordability \cite{cheng2019comparison}. While this makes them an attractive option for short-reach communication lines, implementing DSP in such systems to compensate for nonlinear distortions can undermine their cost-effectiveness and energy efficiency.

\begin{figure}
    \centering
    \includegraphics[width=\linewidth]{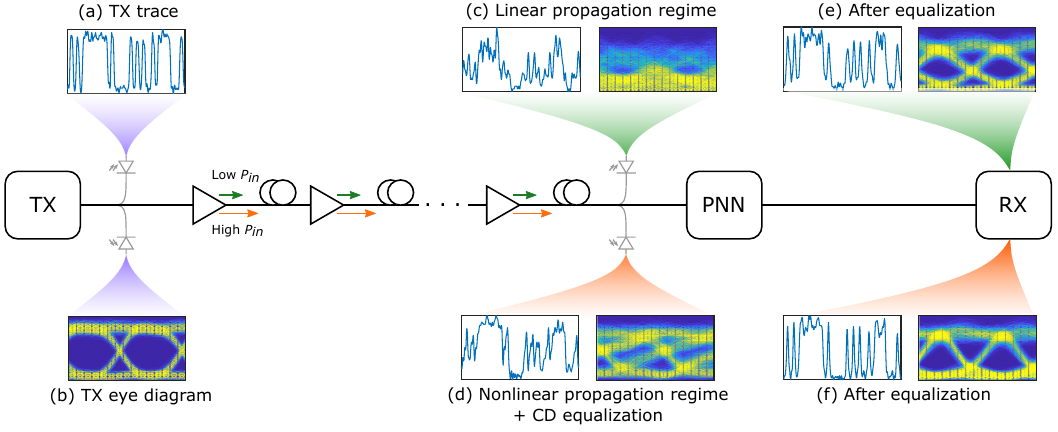}
    \caption{ Schematics of a typical Transmitter/Receiver (TX/RX) optical link featuring a PNN as equalizing unit. The panels highlighted by colored shadows represent the trace and the eye diagram associated with the optical signal in different monitor points along the line. At the TX stage, the optical signal is encoded with a PAM2 modulation (a,b) and propagated via a multi-span link. Each segment features an amplifier (triangle) to restore the input optical power in the fiber $P_{in}$. Panels in the upper half-plane (green shadows) describe the unequalized (c) and equalized (e) signal in linear propagation regime (low $P_{in}$). Panels in the lower half-plane (orange shadows) describe the unequalized (d) and equalized (f) signal in nonlinear propagation regime (high $P_{in}$ and equalized CD). Intensity traces and eye diagrams are actual experimental data.}
    \label{fig:txrx_scheme}
\end{figure}

To address this challenge, we propose a fully optical solution for nonlinear distortion equalization in IMDD systems. Our solution, implemented via a Photonic Neural Network (PNN) \cite{staffoli2025silicon}, aims to eliminate (or at least alleviate) the need for DSP in these systems, significantly reducing costs and energy demands while preserving their simplicity and practicality for short-distance communication. This shift can redefine the economic and technical viability of IMDD systems, unlocking new possibilities for efficient and sustainable telecommunications. \textbf{Figure \ref{fig:txrx_scheme}} presents the schematics of a typical Transmitter/Receiver (TX/RX) optical link with a possible implementation of the PNN device as an equalizer. The modulated optical signal, encoding a 2-level Pulse Amplitude Modulation (PAM2), accumulates distortions while propagating in a multi-span link. The upper half-plane analyzes a linear propagation regime, corresponding to low input power $P_{in}$ in every fiber span, and the transmission degradation is mainly attributable to CD. The lower half-plane analyzes a nonlinear propagation regime, where $P_{in}$ is brought to higher values and CD is equalized, leaving SPM as the primary impairment source. The comparison of the unequalized eye diagram highlights the different nature of effects occurring during propagation in the two regimes. However, the PNN on the RX side restores the eye diagram apertures at the end of the line by operating an optical signal processing on the distorted sequences.

Our approach is based on a photonic integrated circuit implementing a feed-forward PNN. It is trained to recognize and compensate for eye diagram closure, operating the processing directly on the distorted optical signal, minimizing latency. The PNN design features a $N$-lines Finite Impulse Response filter in the optical domain, processing an input optical signal $x(t)$ (complex field) to generate an output $y(t)$ according to
\begin{equation}
    y(t) = \sum_{i=1}^{N} x\left[ t - (i-1)\Delta t \right] a_i k_i \text{e}^{j\phi_i}.
    \label{eq:cperc}
\end{equation}
The input signal $x(t)$ is first split into $N$ separate waveguides (channels, or taps), where each duplicate undergoes a delay that is an integer multiple of $\Delta t$. The delay is imposed by propagating the copies through spirals of different lengths. The channels are numbered from 1 to $N$, the first hosting no spirals and the $N$-th corresponding to the maximum delay. After accumulating the respective delay, each copy is modified by applying a trainable amplitude weight $a_i$ ($0\leq a_i \leq 1$) and a trainable phase shift $\text{e}^{j\phi_i}$. Each channel is then affected by another fixed contribution to amplitude weights $k_i$ introduced by the coupling losses, the propagation losses in the spirals, splitters, combiners, and other fabrication imperfections. The calibration procedure leading to the estimation of the $k_i$ values is reported in Section \ref{sec:experimental_setup}. Once all delayed and weighted copies have been processed, they are recombined and summed together to produce the final output. 

The implemented PNN device layout features $N = 8$ and $\Delta t = 25 $ ps and is optimized to operate on 10 Gbaud signals, providing an optical sampling of 4 samples per symbol and an overall observation window of $\Delta t\times (N-1) = 175$ ps. In a previous application \cite{staffoli2025silicon}, the PNN device was used to counteract the intersymbol interference (ISI) generated by CD in an up-to 125 km long optical link. Dispersive effects in fiber act as a phase mask in time applied to each propagating pulse, as testified by the impulse response associated with fiber propagation:
\begin{equation}
    h_f(t) = e^{-j\frac{1}{2\beta_2L}t^2}.
    \label{eq:h}
\end{equation}
Here, $\beta_2$ is the group velocity dispersion parameter, and $L$ is the fiber length. In this context, the PNN device acts as a tunable Finite Impulse Response (FIR) Filter operating in the optical domain. It resembles a convolution on a finite time interval trained to mimic the inverse of $h_f(t)$ by applying a phase mask only ($a_i = 1$ for every $i$). CD equalization has been proven up to 125 km with a standard single-mode fiber (SM-28) on 10 Gbaud PAM2 and PAM4 signals \cite{staffoli2025silicon,staffoli2023equalization, staffoli2024chromaticPW} and Orthogonal Frequency Division Multiplexed (OFDM) signals \cite{marciano2025chromatic}. 

 In this paper, we test the equalization capabilities of this PNN device on different and more demanding transmission scenarios. These new transmission regimes are made possible by using a fiber loop, which features a Tunable Dispersion Compensating unit (TDC) in the loop, therefore emulating multi-span propagation. In the linear propagation regime, the novel applications regard CD equalization up to 200 km and the equalization of CD residuals. In the nonlinear propagation regime, the novel applications regard SPM equalization. Furthermore, a PNN design for 100 Gbaud applications against SPM and XPM is discussed.

Let us further comment that the PNN device used as a trainable (linear) FIR filter yields CD equalization since its linear nature. In this case, the role of the nonlinear activation function, represented by the square modulus applied at the end-of-line photodetector, remains marginal. On the other hand, for nonlinear equalization, the activation function is of primary importance. In this situation, the PNN device is the linear stage of a time-delayed complex perceptron equipped with the detection process as the activation function \cite{bishop2006pattern}. The PNN device alone cannot compensate for SPM. It is the combined action of the linear and nonlinear stages that counteracts the signal distortions. While applying the FIR filter design to CD equalization seemed straightforward, the application of an all-optical time-delayed complex perceptron \cite{mancinelli2022photonic} to nonlinear equalization appears less obvious.

The paper is structured as follows. Section \ref{sec:fiber_loop} covers the implementation of the fiber loop used for multi-span propagation, the tested transmission scenarios, and the methods for assessing transmission quality. Specifically, the experimental setup designed for 10 Gbaud applications is detailed in Section \ref{sec:fiber_loop_exp}, while the simulated counterpart for 100 Gbaud applications, including dataset generation, is described in Section \ref{sec:fiber_loop_sim}. Results are presented and analyzed beginning with the experimental findings on CD and self-phase modulation (SPM) equalization in Section \ref{sec:results_exp}. Simulated results for SPM and XPM equalization at higher baud rates are discussed in Section \ref{sec:results_sim}. Finally, Section \ref{sec:conclusion} provides the conclusions, followed by supplementary details on the experimental setup in Section \ref{sec:experimental_setup} and its simulated counterpart in Section \ref{sec:simulated_setup}.

\section{Methods}
\label{sec:fiber_loop}
\subsection{Experimental setup for 10 Gbaud propagation}
\label{sec:fiber_loop_exp}

\begin{figure}
    \centering
    \includegraphics[width=\linewidth]{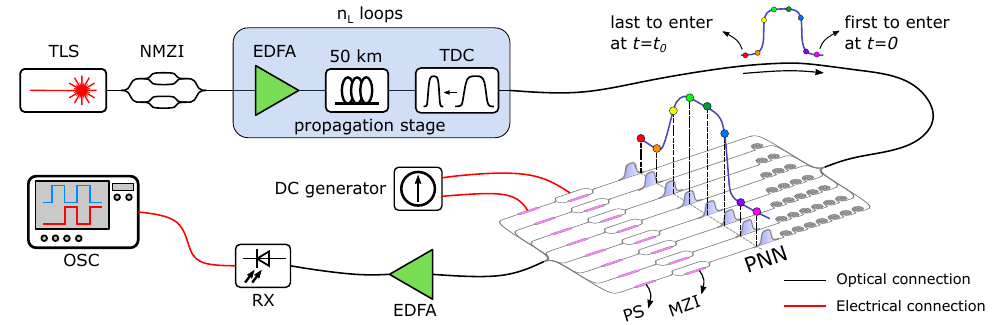}
    \caption{Simplified schematics of the experimental setup. The CW light emitted by a Tunable Laser Source (TLS) is modulated via a Nested Mach-Zehnder Interferometer and then proceeds through the propagation stage. The propagation occurs as multiple passages across an optical fiber loop (blue rectangle) enclosing an Erbium-Doped Fiber Amplifier (EDFA), a 50 km fiber span, and a Tunable Dispersion Compensator (TDC). The distorted signal is optically processed by the PNN, whose amplitude and phase weights are implemented by Mach-Zehnder Interferometers (MZIs) and Phase Shifters (PSs) controlled by a DC generator. A final amplification stage leads the signal to the receiver (RX), and finally, it is acquired by the Oscilloscope (OSC). In the top-right corner, an illustration describes the PNN processing of a 100 ps pulse, with the optical samples (colored dots) being parallelized in time.}
    \label{fig:setup_simplified}
\end{figure}

\textbf{Figure \ref{fig:setup_simplified}} presents a schematized version of the experimental setup (a more detailed description is provided in Section \ref{sec:experimental_setup}). In the transmission stage, a tunable laser source (TLS) operating at 1550 nm is modulated via a Nested Mach-Zehnder Interferometer (NMZI) as a 10 Gbaud PAM2 signal based on a Pseudo Random Binary Sequence (PRBS). The optical sequence then reaches the propagation stage, where the distortions induced by linear and nonlinear effects are accumulated. The propagation stage emulates a multi-span transmission by recirculating the signal through a fiber loop whose elements are enclosed in the blue rectangle in Figure \ref{fig:setup_simplified}. At every loop cycle, light propagates in a 50 km standard SMF-28 fiber span preceded by an Erbium-Doped Fiber Amplifier (EDFA) that restores the input power in fiber up to 9 dBm. A Tunable Dispersion Compensator (TDC) can be used to impose a dispersion $d_{TDC} \in [-900,0]$ ps nm$^{-1}$ to remove the accumulated CD. Each propagation is then made of an arbitrary number of loop cycles $n_L$, namely multiple passages across the elements enclosed in the blue rectangle. This corresponds to a total propagation distance of $L = n_L \times 50$ km, during which the distortions accumulate in the optical signals. After the loop, the signal proceeds to the PNN device, which uses Mach-Zehnder Interferometers (MZI) and Phase Shifters (PS) to provide the amplitude and phase weights, respectively. These are driven by the thermo-optic effect generated by electrical currents flowing through micro-heaters placed on top of the waveguides. A DC generator controls the 15 PNN parameters constituted by 8 amplitude and 7 phase weights, with the micro-heater driving the phase of the first channel left unconnected and chosen as a reference.

Figure \ref{fig:setup_simplified} also illustrates the working principle of the PNN device upon the arrival of a 100 ps optical pulse, which is pictured in the top-right corner. Colored dots equally spaced by $\Delta t = 25$ ps individuate some optical samples in the input pulse, with the purple one being the first entering the PNN device at $t=0$ and the red one being the last at $t = t_0 = 175$ ps. Once the pulse has entered the PNN device, it is split along 8 lanes, where the 8 copies travel in parallel, accumulating a relative delay of $\Delta t$. This results in the colored optical samples associated with the different copies of the input signal being temporally aligned after the delay stage at time $t = t_0$. The optical samples serially inserted in the PNN device are now parallel propagating in the corresponding channels, finally reaching the summation stage at $t=t_0$. In other words, the delay lines parallelize the input information, allowing for simultaneous processing of pieces of optical information separated initially in time. 

After the optical processing provided by the PNN device, a final amplification stage compensates for its losses and sends the signal to the receiver stage (RX). The oscilloscope (OSC) limits the overall bandwidth of the apparatus to 16 GHz and a sampling frequency of 80 GSa s$^{-1}$. A single oscilloscope acquisition occupies a 1 $\mu$s window, resulting in a $10^4$ symbols sequence. 

\subsubsection{Methods and explored transmission scenarios}
The PNN equalization capabilities are tested in different transmission scenarios determined by the number of loops $n_L = 0,...,9$, input power in fiber $P_{in} = \{0,\,3,\,6,\,9\}$ dBm at every recirculation, and the dispersion $d_{TDC} = \{- 900,\, 0\}$ ps nm$^{-1}$ imposed by the TDC. In particular, we can distinguish between two different regimes explored during the experiments. In the linear propagation regime, the input power is set to $P_{in} = 0$ dBm, and the PNN is called to compensate for CD ($d_{TDC} = 0$ ps nm$^{-1}$) or its residuals ($d_{TDC} = -900$ ps nm$^{-1}$). On the other hand, in the nonlinear propagation regime, $P_{in} > 0$ dBm and $d_{TDC} = -900$ ps nm$^{-1}$, and the PNN is called to equalize SPM-related distortions. Unfortunately, XPM observation was prevented by setup limitations, including the limited available bandwidth and a minimum achievable detuning of 200 GHz between two available laser sources, corresponding to a too-short walk-off distance \cite{hui1999cross}. 

The transmission quality is accessed via a loss function, whose input parameters are represented by the number of loops $n_L$ and the array of currents driving the PNN weights. A loss function call propagates the optical pulses for the assigned number of loops, sets the current to the PNN, and acquires the corresponding RX trace. The outputs are the Bit Error Rate (BER), and the result of the 2-sample Separation loss function (i.e., the separation of the optical levels) presented in \cite{staffoli2025silicon}. This analog function is used for PNN training via the Particle Swarm Optimizer (PSO) to maximize the eye diagram aperture. The unequalized and equalized transmissions are compared via the corresponding BER versus Power at the receiver ($P_{RX}$) profiles. A scan over $P_{RX}$ is performed via a Variable Optical Attenuator (VOA) placed in front of the end-of-line receiver (VOA 3 in \textbf{Figure \ref{fig:setup}}), and for each $P_{RX}$, the BER is evaluated as the average of 50 repeated measurements, resulting in a minimum measurable value of $1/(50 \times 10^4) = 2\times 10^{-6}$. 

In the experimental context, all the measurements maintain the same periodic input sequence represented by a $2^{10}$ bit long PRBS of order 10. Notice that each oscilloscope acquisition allows the observation of at least 9 periods of the mentioned sequence, and the periodicity is exploited to align the acquired trace with the corresponding target sequence by their cross-correlation. The proposed approach has no distinction between the training and testing data set since the chosen input sequence allows the PNN to experience all the observable distortion conditions generated by adjacent symbols. Indeed, the pulse time broadening generated by CD is linked to the group velocity dispersion parameter $\beta_2 = -0.022$ ps$^2$ m$^{-1}$, the fiber length $L$, and the transmission bandwidth $\Delta \omega = 2\pi \times 5$ GHz by
\begin{equation}
    \Delta T = \beta_2 L \Delta \omega.
    \label{eq:bit_broadening}
\end{equation}
This measures how much information about a symbol has spread in the adjacent ones, and this is strongly dependent on the propagation conditions. The chosen worst-case scenario in the linear propagation regime is represented by $L = 200$ km and $d_{TDC} = 0$ ps nm$^{-1}$, leading to $\Delta T = 140$ ps and a total symbol time width of $1/B+ \Delta T = 240$ ps, with $B$ being the baud rate. The generated ISI involves 3 adjacent symbols (the central one and its left and right prime closest). This represents a limit case for the PNN since the total bit width is more significant than the maximum time window of $(N-1)\times \Delta t = 175$ ps that the PNN can parallelize and process simultaneously. On the other hand, in the nonlinear propagation regime where $d_{TDC} = -900$ ps nm$^{-1}$, CD acts within every 50 km segment before being equalized. At each loop, the broadening reaches $\Delta T = 35$ ps and is subsequently counteracted by the TDC. Therefore, in both propagation regimes, the distortions result from the interaction of at most 3 adjacent bauds, making the chosen PRBS input sequence sufficient to let the PNN explore all the possible distortions.

\subsection{Simulated setup for 100 Gbaud propagation}
\label{sec:fiber_loop_sim}
The limitations of the experimental setup have been circumnavigated by implementing a simulated version of the optical line. This allowed overcoming technical challenges and performing a study on the scalability of the proposed approach to different transmission scenarios, including 100 Gbaud modulation and XPM-induced distortions. A detailed description of the code and the parameters used in the simulation is presented in Section \ref{sec:simulated_setup}. 

The simulated setup is implemented via MATLAB software, representing a simplified version of the experimental line described in the previous section. It features two laser sources, Probe and Pump, separated by a detuning $\Delta \nu = 50$ GHz and independently modulated via NMZIs. The so-obtained optical signals are then sent to the loop for multi-span propagation. Inside the loop, the fiber propagation through the 50 km span is simulated via the Split-Step Fourier Method (SSFM) \cite{agrawal2013nonlinearSSFM}, subsequently removing the CD contribution to mimic the action of the TDC. The Optical Signal-To-Noise (OSNR) degradation introduced by the multiple passages through the EDFA is accounted for by following the approach proposed in \cite{agrawal2021fiber,chomycz2009osnr}. The simulated setup includes a parametric version of the PNN device, with tunable unitary delay $\Delta t$ and number of taps $N$. The processed optical signal is then sent to the end-of-line photodetector, whose noise generation is modeled as the sum of thermal and shot noise contributions.

\subsubsection{Methods and explored transmission scenarios}
The experimental results obtained with 10 Gbaud signals in terms of BER versus PRX profiles and eye diagram apertures served as a benchmark for code calibration. After that, the bandwidth of each instrument (NMZI, RX, OSC, ...) is properly rescaled to accommodate a 100 Gbaud modulation. Each transmission scenario is identified by the number of loops $n_L = 0,...,9$, the probe input power in fiber $P_{pr} = [0,10]$ dBm, and the pump input power in fiber $P_{pu} = [-30,10]$ dBm, with $P_{pu} = -30$ dBm standing for the pump being off. Each measure accessing the transmission quality returns the BER and the 2-sample separation loss function, which is provided to the PSO for PNN training. The loss function receives as input an optical sequence (complex field) to be processed, the associated target sequence, and the PNN weights $a_{i} = [0,1]$ and $\phi_{i} = [0, 2\pi)$. 

\begin{figure}
    \centering
    \includegraphics[width=\linewidth]{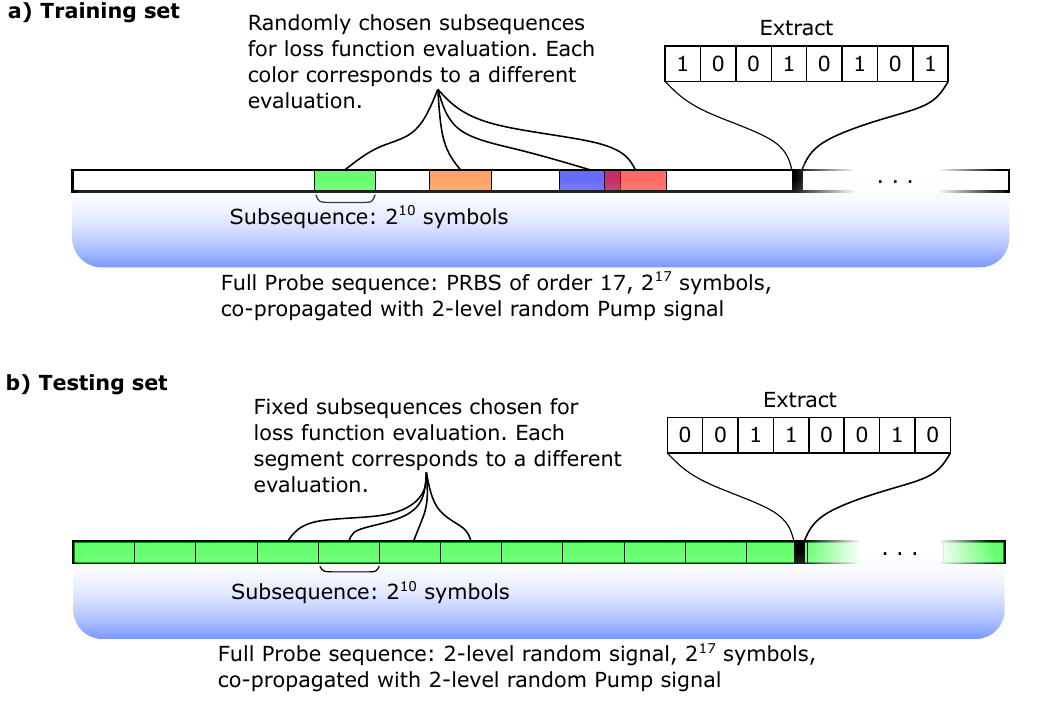}
    \caption{Illustration of training (a) and testing (b) datasets creation used in simulations at 100 Gbaud. (a) For the training set, the probe sequence encodes a $2^{17}$ symbols PRBS sequence of order 17 co-propagated with a 2-level random pump sequence. The propagation is repeated for different configurations of $P_{pr}$, $P_{pu}$, and $n_L$, obtaining the data set specific to each scenario. Each loss function evaluation during the training is operated on a sub-sequence of $2^{10}$ symbols (colored segments) derived from the main one by randomly choosing a starting point and selecting the subsequent $2^{10}$ symbols. (b) For the testing set, both the probe and pump signals encode 2-level random sequences. Again, the propagation is repeated for different configurations of $P_{pr}$, $P_{pu}$, and $n_L$, obtaining the data set specific to each scenario. The resulting probe trace is divided $\sim100$ consecutive subsequences of $2^{10}$ symbols (green segments), each separately used for loss function evaluation.}
    \label{fig:datasets_explanation}
\end{figure}

The approach proposed with a 10 Gbaud modulation, which always uses the same $2^{10}$ bit long PRBS sequence for every measure, no longer applies to higher frequencies. Indeed, for a 100 Gbaud signal propagating in a 50 km link, Equation \ref{eq:bit_broadening} estimates a pulse time-broadening of around 350 ps, corresponding to 35 times the original baud time width. It would require the propagation of an incredibly long sequence to allow the PNN to explore all the possible distortion conditions generated by the interference of 35 symbols in a single measure. The unfeasability of this solution necessitates reintroducing the distinction between training and testing data sets, whose generation is depicted in \textbf{Figure \ref{fig:datasets_explanation}}. The training set in panel (a) is based on a $2^{17}$ symbols long PRBS sequence of order 17 co-propagated with a 2-level random pump sequence. The same two sequences are propagated in every transmission configuration regarding $P_{pr}$, $P_{pu}$, and $n_L$, obtaining the data set specific to each scenario. Subsequently, each loss function evaluation during the training is operated on a sub-sequence of $2^{10}$ symbols derived from the main one by randomly choosing a starting point and selecting the subsequent $2^{10}$ symbols. Similarly, the generation of the testing data set in panel (b) originates from the propagation of the probe and pump signals, this time both encoding a 2-level random sequence. The resulting probe trace is divided $\sim100$ consecutive subsequences of $2^{10}$ symbols, each separately used for loss function evaluation. The testing procedure outputs a BER versus $P_{RX}$ profile where each point is obtained as the average loss function evaluation over the mentioned 100 subsequences.

\section{Results}
\subsection{Experimental tests}
\label{sec:results_exp}
\subsubsection{Linear regime}
\begin{figure}
    \centering
    \includegraphics[width=\linewidth]{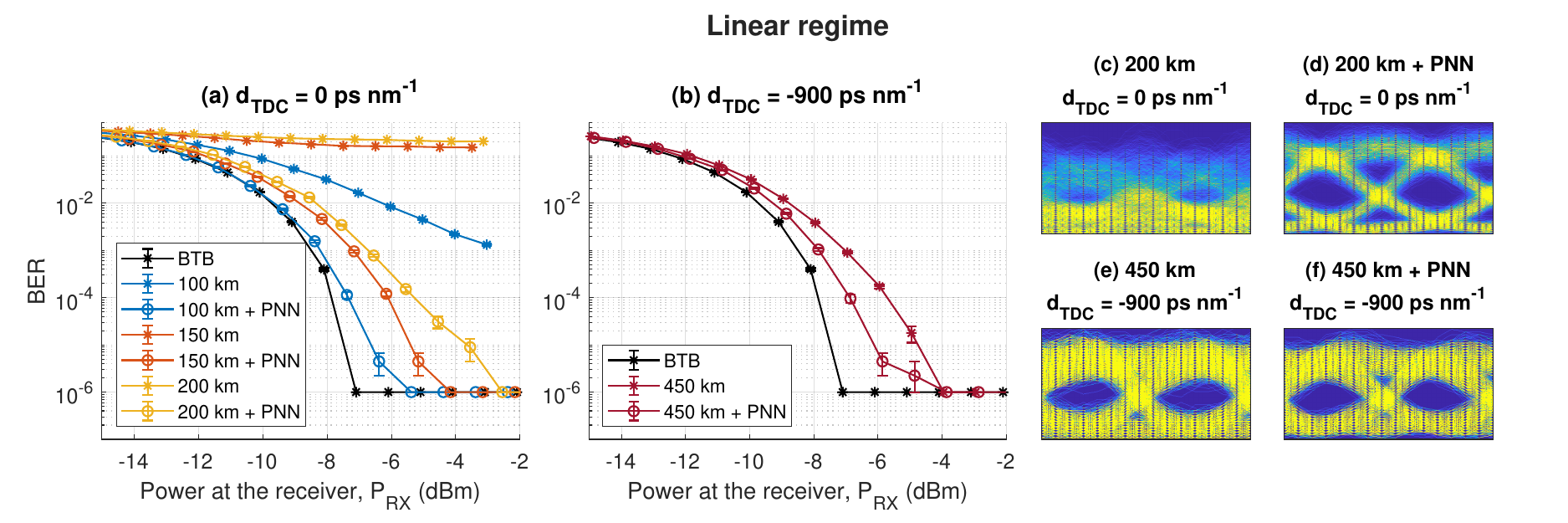}
    \caption{Experimental results for equalization at 10 Gbaud in linear propagation regime ($P_{in} = 0$ dBm). (a) BER versus $P_{RX}$ profiles obtained with $d_{TDC} = 0$ ps nm$^{-1}$ for unequalized (stars) and equalized (circles) transmission. Different colors in the lines indicate different propagation lengths $L$. (b) BER versus $P_{RX}$ profiles obtained with $d_{TDC} = -900$ ps nm$^{-1}$ for unequalized (stars) and equalized (circles) transmission for $L=450$ km. In (a) and (b), the curve generated for Back-To-Back transmission (black stars) sets the reference performance. Error bars on each point correspond to a 68\% confidence interval estimated by bootstrap method with $10^3$ iterations \cite{efron1992bootstrap,ader2008advising}. Null BER points have been replaced with the minimum measurable value, which also sets a limit for the lower bound of the confidence intervals. (c,d) Eye diagrams for the unequalized (c) and equalized (d) transmission in $L=200$ km and $d_{TDC} = 0$ ps nm$^{-1}$. (e,f) Eye diagrams for the unequalized (e) and equalized (f) transmission in $L=450$ km and $d_{TDC} = -900$ ps nm$^{-1}$.}
    \label{fig:lin_ber}
\end{figure}

\textbf{Figure \ref{fig:lin_ber}} reports the results obtained in the linear propagation regime, featuring $P_{in} = 0$ dBm at every loop. The BER versus $P_{RX}$ profiles (panel (a)) are obtained setting $d_{TDC} = 0$ ps nm$^{-1}$, so the PNN is called to compensate for CD. The trainings are performed by setting the amplitude weights to $a_{i} = 1$ for $i=1,\dots,8$ and optimizing the phases to mimic the impulse response of Equation \ref{eq:h}. The so-obtained results extend those shown in \cite{staffoli2024chromaticPW} since the propagation distance has increased over 125 km, up to 200 km. The unequalized BER profiles (colored stars) depart from the Back-To-Back curve (black stars), with the ISI generated by CD becoming increasingly severe with the propagation distance. The trained PNN lowers the BER profiles, restoring the eye diagram aperture from panel (c) to panel (d). On the other hand, the results reported in panel (b) are obtained by setting $d_{TDC} = -900$ ps nm$^{-1}$, removing the most contributions of CD at every loop. The departure of the BER versus $P_{RX}$ profile (dark red stars) from the BTB curve (black stars) becomes evident only after long propagation due to CD residual accumulation, which the trained PNN mitigates (dark red circles). This improvement also appears in the corresponding eye diagrams in panels (e) and (f).

\subsubsection{Nonlinear regime}
\begin{figure}
    \centering
    \includegraphics[width=\linewidth]{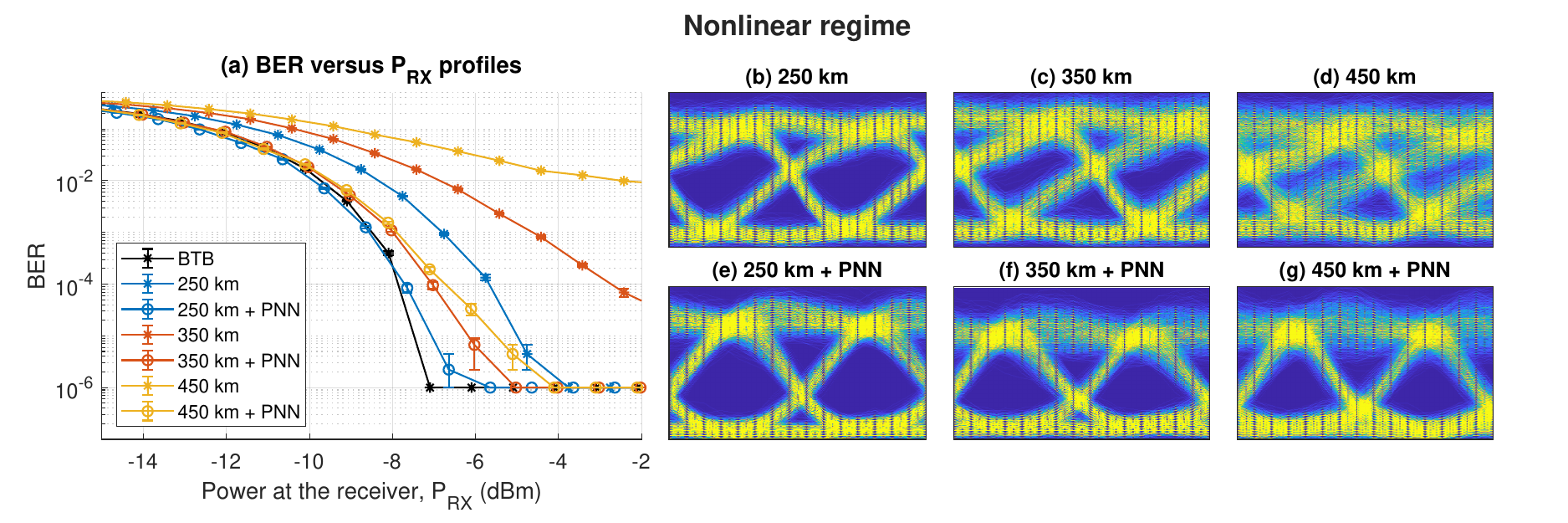}
    \caption{Experimental results for equalization at 10 Gbaud in nonlinear propagation regime ($P_{in} = 9$ dBm, worst case scenario). (a) BER versus $P_{RX}$ profiles obtained with $d_{TDC} = -900$ ps nm$^{-1}$ for unequalized (stars) and equalized (circles) transmission. Different colors in the lines indicate different propagation lengths $L$. The curve generated for Back-To-Back transmission (black stars) sets the reference performance. Error bars on each point correspond to a 68\% confidence interval estimated by bootstrap method with $10^3$ iterations \cite{efron1992bootstrap,ader2008advising}. Null BER points have been replaced with the minimum measurable value, which also sets a limit for the lower bound of the confidence intervals. (b-g) Eye diagrams for the unequalized (b-d) and equalized (e-g) transmission in (b,e) $L=250$ km, (c,f) 350 km, and (d,g) 450 km. }
    \label{fig:nl_ber}
\end{figure}

In correspondence with a higher input power in fiber ($P_{in} > 0$ dBm), the PNN is trained to equalize SPM-induced distortions, which have been isolated from CD contributions by setting $d_{TDC} = -900$ ps nm$^{-1}$. The impact of nonlinear effects on the transmission quality is presented in \textbf{Figure \ref{fig:nl_ber}}, obtained in the worst-case scenario of $P_{in} = 9$ dBm. The accumulation of nonlinear distortions with the propagation distance appears in panel (a) as an increasing gap between the unequalized BER versus $P_{RX}$ profiles (colored stars) and the BTB performance (black stars). This is reflected in panels (b-d) as a progressive tilting of the eye diagram. Again, the PNN beneficial action restores the transmission's quality, lowering the BER profiles (colored circles) closer to the BTB curve and opening the eye diagrams (panels (e-g)). 

\begin{figure}[ht!]
    \centering
    \includegraphics[width=\linewidth]{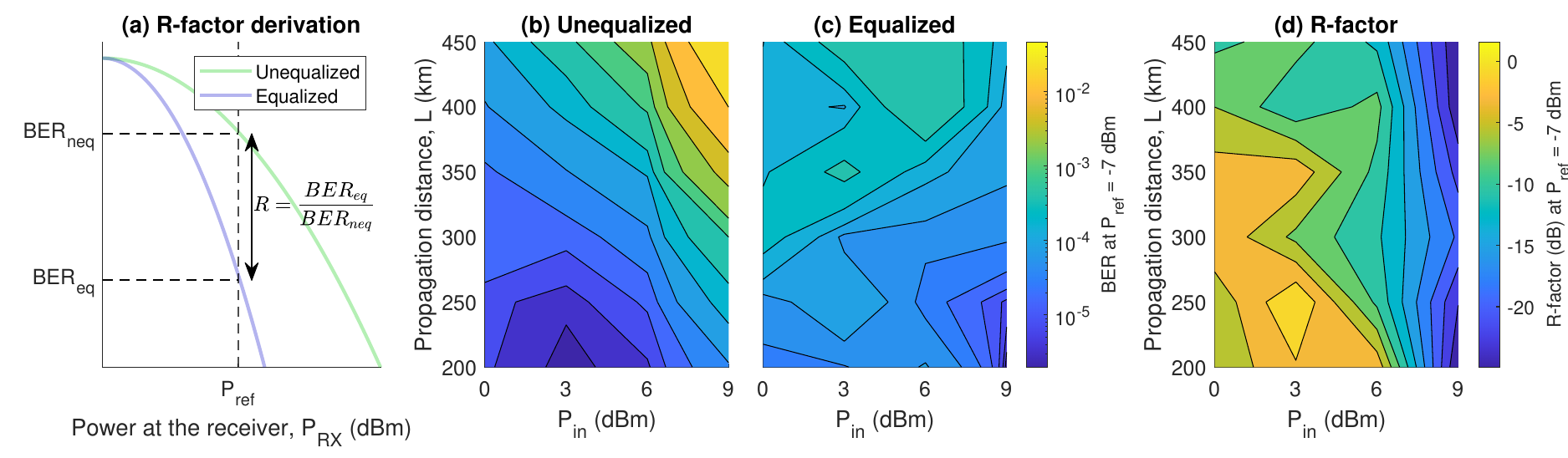}
    \caption{(a) Illustration of the BER reduction factor (R-factor) derivation at a given reference power $P_{ref}$ from two unequalized (green) and equalized (blue) BER versus $P_{RX}$ curves. (b) Unequalized and (c) equalized BER measured at $P_{ref} = -7$ dBm mapped as a function of the input power in fiber $P_{in}$ and the total propagation distance $L$. Null BER points have been replaced with the minimum measurable value. (d) R-factor map at $P_{ref} = -7$ dBm as a function of $P_{in}$ and the propagation distance $L$. }
    \label{fig:nl_maps}
\end{figure}

To quantify the figure of merit brought by the PNN, we introduce the BER reduction factor $R$, whose derivation is illustrated in \textbf{Figure \ref{fig:nl_maps}}(a). Considering the unequalized BER versus $P_{RX}$ profile obtained in a given transmission scenario, $BER_{neq}$ represents the linearly interpolated value at a reference power at the receiver $P_{ref}$. Analogously, $BER_{eq}$ represents its homologous obtained from the equalized profile generated after the introduction of the PNN. The $R$-factor results from the ratio (expressed in dB) of the two as $R=BER_{eq}/BER_{neq}$, with lower values representing better equalization performance of the PNN. Notice that the choice of $P_{ref}$ value aims to produce informative BER maps with meaningful variations between equalized and unequalized BERs. Taking Figure \ref{fig:nl_ber}(a) as a reference, a too-low $P_{ref}$ explores the left-most region of a typical BER versus $P_{RX}$ curve where noise is the dominant contribution. On the other hand, a too-high $P_{ref}$ may result in low BER estimates corresponding to the statistical limit. Thus, the tradeoff to obtain informative maps from a family of curves is found with $P_{ref}$ that intercepts the last non-null (i.e., higher than the statistical limit) BER value in the lowest curve. 

Figure \ref{fig:nl_maps}(b) reports the $BER_{neq}$ map for $P_{ref} = -7$ dBm as a function of $P_{in}$ and $L$. The $BER_{neq}$ values increase along both axes, and the worst-case scenario is identified by the yellow region in the top-right angle, corresponding to the combination of the highest input power and longest propagation distance. The introduction of the PNN makes the $BER_{eq}$ map of Figure \ref{fig:nl_maps}(c) more uniform, eliminating the presence of the peak in the top-right angle. Overall, the PNN leads to an improved transmission quality, also testified by the $R$-factor map of Figure \ref{fig:nl_maps}(d). Configurations in which $R>0$ are negligible since the corresponding $BER_{neq}$ is far below the pre-FEC threshold of $10^{-3}$.

\subsubsection{Weights analysis}
\begin{figure}
    \centering
    \includegraphics[width=\linewidth]{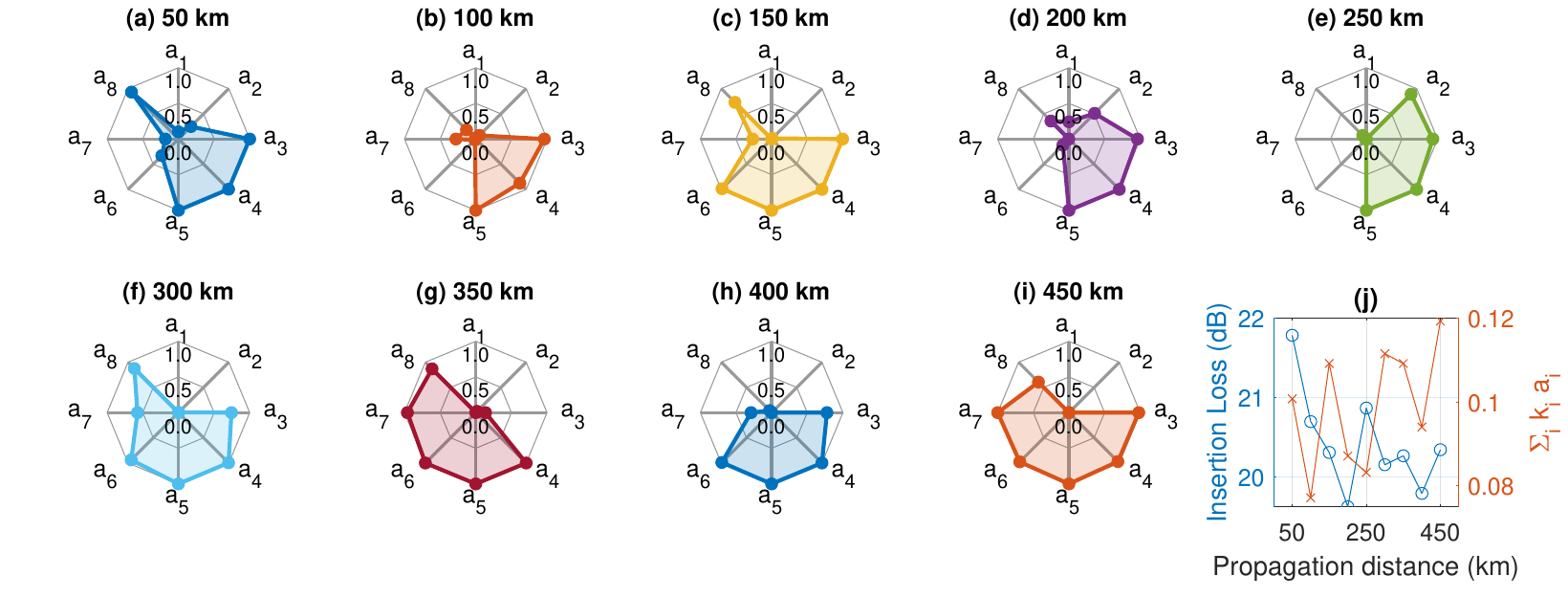}
    \caption{(a-i) Optimal amplitude weights $a_i$ after PNN training obtained for channel equalization in nonlinear regime ($P_{in} = 9$ dBm) at different propagation distances. (j) Measured insertion loss (left axis) and overall channel apertures obtained as $\sum_{i=1}^8 k_i a_i$ (right axis) for the PNN device while operating (i.e., with active optimal currents) signal equalization in different propagation distances.}
    \label{fig:ch_apertures}
\end{figure}

\textbf{Figure \ref{fig:ch_apertures}} presents the optimal amplitude weights (a-i), the corresponding overall channel apertures ($\sum_{i=1}^8 k_i a_i$) and insertion loss (IL) (j) obtained for $P_{in} = 9$ dBm at different propagation distances. We expected to see a trend where, as $L$ increases, the PNN device features more adjacent channels in the open state (i.e., $a_i \simeq 1$), resulting in a partial reduction of the IL. Indeed, longer $L$ allows more SPM-related distortions to accumulate and possibly interact with residual CD. This increases the task's complexity, making it necessary to combine more and more pieces of optical information from different time instants. However, many factors influence the training procedures (noise, polarization rotation during propagation, PNN device misalignment, etc.), making the expected trend less evident from the measured data. Support for our expectations comes from simulations, which emulates the same experimental conditions as those associated with Figure \ref{fig:ch_apertures}. Multiple training sessions for SPM equalization are performed, imposing different constraints on the number $n_{op}$ of adjacent channels, which are set in the open state while the others are closed. It is found that a condition with $n_{op} = 3$ provides equalization for $ L = 100$ km, in accordance with Figure \ref{fig:ch_apertures}(b). For $L=450$ km, an increased value of $n_{op}$ is needed to equalize the transmission with the same performance.

\begin{figure}
    \centering
    \includegraphics[width=\linewidth]{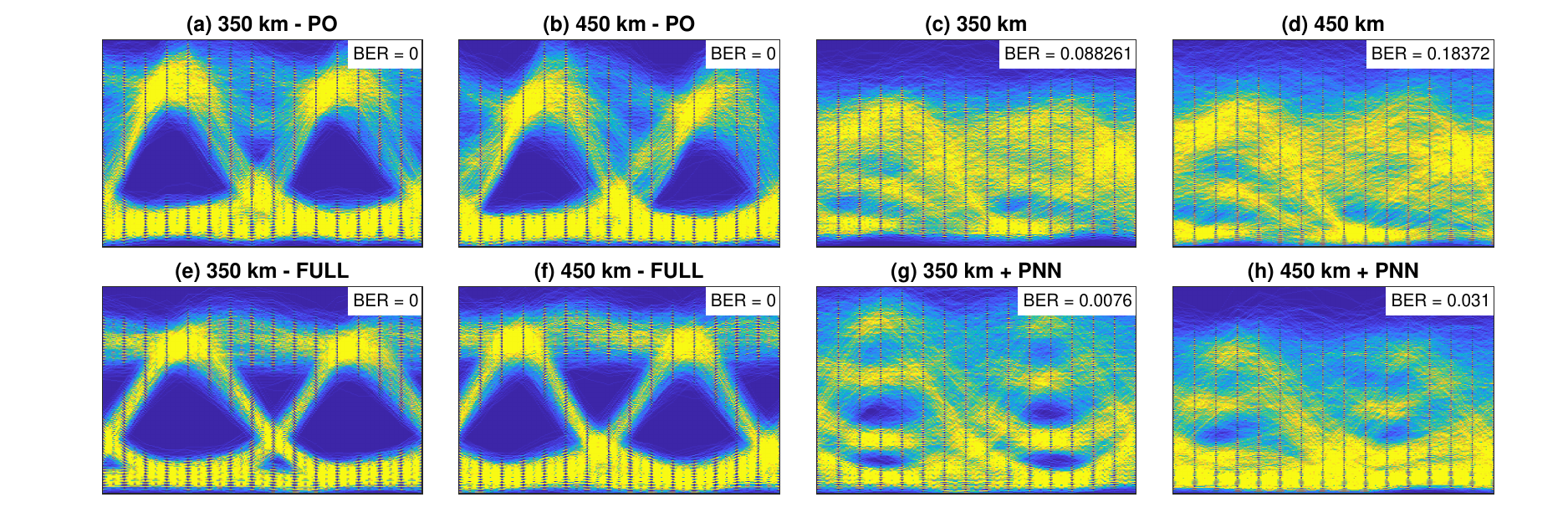}
    \caption{Eye diagrams at the receiver after signal propagation at $P_{in} = 9$ dBm. BER values reported in each panel are measured at $P_{RX} \sim 0$ dBm. (a-b) Equalized eye diagrams after propagation in (a) 350 km and (b) 450 km, with the PNN trained in Phase-Only (PO) configuration. (e-f) Equalized eye diagrams after propagation in (e) 350 km and (f) 450 km, with the PNN trained in Full configuration. (c,d,g,h) Unequalized (c,d) and equalized (g,h) eye diagrams for 10 Gbaud PAM4 modulation in (c,g) 350 km and (d,h) 450 km. }
    \label{fig:eyes_po_pam4}
\end{figure}

Tuning the amplitude weights makes the training time longer, but it becomes fundamental in the attempt to equalize nonlinear effects. A training performed in Phase Only (PO) configuration, namely fixing $a_i = 1$ for every $i$ and adjusting the phase weights solely, produces a sub-optimal equalization compared to training performed in FULL configuration, where both amplitude and phase weights can be adjusted. \textbf{Figure \ref{fig:eyes_po_pam4}}(a-b) shows the eye diagrams after 350 km and 450 km propagation equalized with the PNN in PO configuration. Despite the BER being null, distortions and ripples appear, causing a broadening of the high level and the transient lines. On the contrary, the equalized eye diagrams obtained in FULL configuration appear wider and symmetric to the vertical axis, as shown in Figure \ref{fig:eyes_po_pam4}(e-f).

Finally, the eye diagrams for unequalized and equalized PAM4 modulation for $P_{in} = 9$ dBm are reported in Figure \ref{fig:eyes_po_pam4}(c-d) and (g-h), respectively. Transmission is limited by the high noise level introduced by RF amplifiers already at the transmitter stage, which generates non-null BER even in BTB configuration. However, despite the low SNR, the PNN provides up to an order of magnitude of BER reduction, proving its potential in SPM equalization even with multi-level modulation formats.


\subsection{Simulated tests}
\label{sec:results_sim}
\begin{figure}
    \centering
    \includegraphics[width=\linewidth]{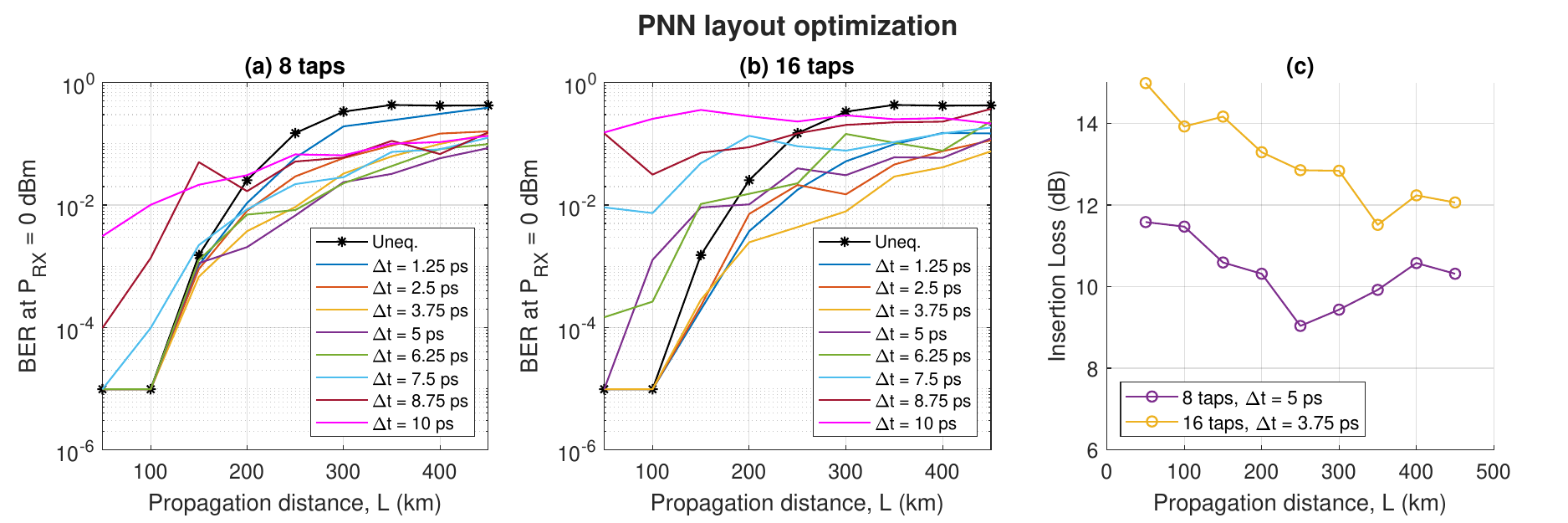}
    \caption{Simulated results for PNN layout optimization to operate with 100 Gbaud signals. (a-b) Equalized BER measured at $P_{RX} = 0$ dBm as a function of the propagation distance $L$. The tested PNN layouts feature (a) $N = 8$ and (b) $N = 16$, respectively. In both cases, a scan over the unitary delay parameter $\Delta t$ is performed, producing the results reported in each colored curve. Null BER points have been replaced with the minimum measurable value. Error bars are not plotted for the sake of clarity. (c) Insertion Loss as a function of the propagation distance measured for the optimized PNN device in operation. The two curves are referred to a PNN featuring $N = 8$, $\Delta t = 5$ ps, and $N = 16$, $\Delta t = 3.75$ ps, namely the PNN configurations offering the best performances in panels (a) and (b). }
    \label{fig:layout_opt}
\end{figure}

In the simulated environment, the bitrate is increased to 100 Gbps to analyze the scalability of the PNN approach to a real-case scenario. A first study is prone to optimize the PNN layout in terms of the number of taps $N$ and unitary delay $\Delta t$ to compensate for SPM in the worst-case scenario, thus fixing the pump input power in fiber to $P_{pu} = 10$ dBm. \textbf{Figure \ref{fig:layout_opt}}(a) presents the performance comparison of an 8-tap layout with various unitary delays. Each curve, corresponding to a different $\Delta t$, reports the BER at $P_{RX} = 0$ dBm obtained after the PNN training for different propagation distances. The scan is performed over $\Delta t$ values ranging from 1/8 to a full baud time-width. The reference performance is set by the unequalized curve (black stars). Similarly, the results of Figure \ref{fig:layout_opt}(b) are referred to the optimization of a 16-tap PNN. Other cases are not an object of this study since $N = 4$ would have represented a too-simplistic layout, while $N > 16$ would drastically increase the insertion losses. 

Considering the 8-tap layout, in panel (a) it appears that a PNN device featuring a large unitary delay $\Delta t \geq 8.75$ ps generates a higher BER compared to the unequalized transmission. The same condition is present in a 16-tap design for $\Delta t \geq 5$ ps up to $L \leq 150$ km, extending to $L=200$ km for $\Delta t \geq 7.5$ ps. This behavior is justified by considering that for long propagation distances, the interplay between SPM and CD (that is, the origin of the distortions) in each 50 km span accumulates, and a mutual influence between adjacent symbols is enforced. The equalization process requires, therefore, more memory, namely, the simultaneous access to an increasing number of consecutive symbols. On the contrary, for short propagation distances, mutual interference is minimal, and equalization requires a limited amount of memory. In our case, PNN devices featuring small $\Delta t$ result particularly suitable for low-memory tasks: indeed, they allow maintaining a small time window for recombination (i.e., memory) while having multiple channels in the open state, and thus containing the insertion losses. In other words, they can ensure high transmission, and thus a high SNR at the receiver, and still avoid bringing to recombination unnecessary pieces of information which would degrade the equalization performance. On the other hand, this optimal condition cannot be achieved for PNN devices featuring larger $\Delta t$. Indeed, having more channels in the open state reduces the insertion losses but inevitably introduces unwanted optical samples in the recombination process that cause distortions rather than eliminating them.

Moving back to the $\Delta t$ optimization process, in both panels (a) and (b) the best performance is identified by the lowest curve, revealing the optimal unitary delays to be $\Delta t= 5 $ ps for $N = 8$ and $\Delta t = 3.75 $ ps for $N = 16$. Despite the lower BER values provided by the optimized 16-taps layout, its better performance is obscured by the higher insertion loss (Figure \ref{fig:layout_opt}(c)) and the more significant number of parameters that result in longer training times and higher chances of training failure. These critical aspects do not justify the slightly improved BER reduction compared to the 8-tap design, which remains the preferred option. 

\begin{figure}[h!]
    \centering
    \includegraphics[width=\linewidth]{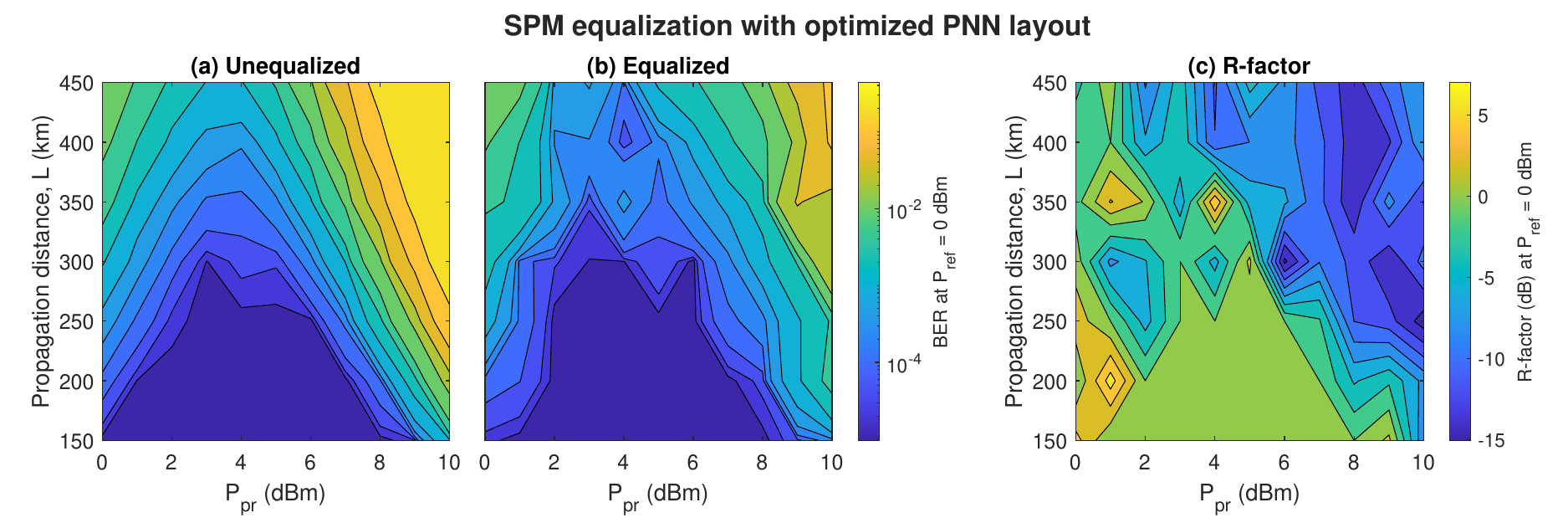}
    \caption{Simulated results for 100 Gbaud PAM2 modulation with an optimized PNN featuring $N = 8$ and $\Delta t = 5$ ps. (a) Unequalized and (b) equalized BER values and (c) corresponding $R$-factor measured at $P_{ref} = 0$ dBm as a function of the probe input power in fiber $P_{pr}$ and the propagation distance $L$. Null BER points have been replaced with the minimum measurable value. }
    \label{fig:spm_maps_sim}
\end{figure}

The optimal 8-tap layout is then tested in various configurations, training the PNN for different combinations of probe input power $P_{pr}$ and propagation distance. The results are reported in \textbf{Figure \ref{fig:spm_maps_sim}} in terms of unequalized and equalized BER at $P_{ref} = 0$ dBm in panels (a) and (b), and the corresponding BER reduction factor in panel (c). The BER values in both the maps (panels (a) and (b)) show a general trend where the isolines describe a reversed U-shape: for fixed $L$, the BER decreases up to $P_{pr} = 3$ dBm, and then the trend is inverted for higher $P_{pr}$. The shape results from the interplay between the OSNR and the SPM effect: as the input power in fiber increases, the OSNR increases, too, and in turn, the BER diminishes. However, for $P_{pr} > 3 $ dBm, the distortions induced by SPM become predominant over the augmented OSNR, causing a BER raise. The highest values are reached in the top-right corner of panel (a), corresponding to the unequalized transmission at high input power and long propagation distance. The trained PNN partially equalizes the SPM-induced distortions, imposing a BER reduction of up to 15 dB. In some cases, the PNN presence can even become detrimental rather than beneficial, as testified by positive $R$-factor values in panel (c) (yellow spots). Sub-optimal training processes may induce these, or, for low $P_{pr}$, SPM-related distortions are quasi-absent, thus not requiring equalization. In this regime, the PNN fails to transmit the signal without modifications, and the optical processing increases the BER.

\begin{figure}
    \centering
    \includegraphics[width=\linewidth]{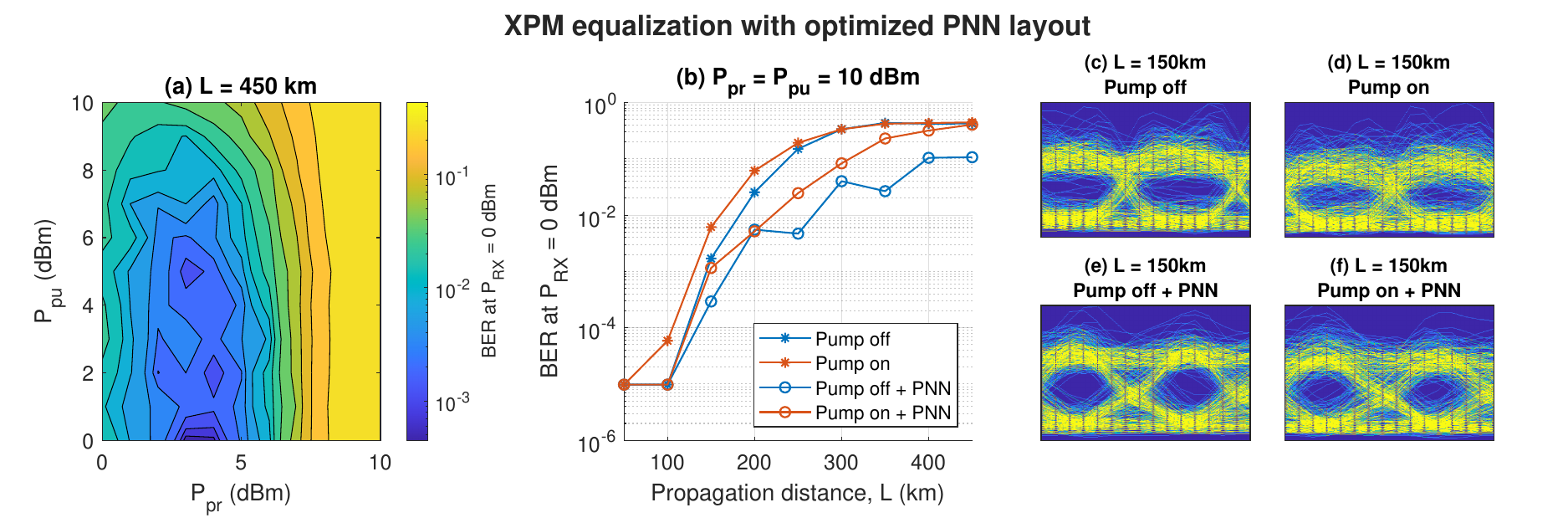}
    \caption{ Simulated results for XPM equalization in a 100 Gbaud PAM2 signal. (a) Simulated BER measured at $P_{RX} = 0$ dBm for $L=450$ km as a function of the probe $P_{pr}$ and pump $P_{pu}$ input power in fiber. (b) Unequalized (stars) and equalized (circles) BER at $P_{RX} = 0$ dBm as a function of the propagation distance for $P_{pr} = 10$ dBm. Blue curves are obtained with the pump signal turned off, while orange curves are obtained by co-propagating the probe with a pump signal at $P_{pu} = 10$ dBm. Error bars are not plotted for the sake of clarity in data visualization. (c-f) Eye diagrams associated with the four profiles presented in (b) for $L = 150$ km.}
    \label{fig:xpm_maps_sim}
\end{figure}

Finally, the same optimized PNN layout has been applied to XPM equalization. The training process relies solely on the probe beam’s initial bit sequence, without accounting for the random power fluctuations in the pump beam, which introduce distortions on the probe beam. As a result, the model is challenged to correct for these unpredictable variations. The first exploratory phase individuates the threshold value for the pump input power in fiber $P_{pu}$ that triggers XPM. Two random sequences of $2^{10}$ bits have been co-propagated up to 450 km in different conditions of $P_{pr}$ and $P_{pu}$, and the corresponding BER at $P_{RX} = 0$ dBm are reported in \textbf{ Figure \ref{fig:xpm_maps_sim}}(a). For fixed $P_{pr}$, the mapped values assume a dependence on the pump power for $P_{pu} > 8$ dBm when the transmission quality is affected by XPM, with a consequent BER increase. This exploratory mapping has been performed by propagating relatively short sequences since using the long-traces approach described in Figure \ref{fig:datasets_explanation} would have been extremely time-consuming. Switching back to the long-traces approach for the following simulations, the PNN is trained to equalize the transmission for $P_{pr} = P_{pu} = 10$ dBm; an interplay between SPM and XPM generates the accumulated distortions. Figure \ref{fig:xpm_maps_sim}(b) presents the unequalized (stars) and equalized (circles) BER values at $P_{RX} = 0$ dBm. Even for short propagation distances, the presence of the pump signal (orange stars, pump on) worsens the transmission compared to the probe-only propagation (blue stars, pump off). The trained PNN manages to improve the transmission quality even in the presence of the XPM effect despite being unable, in general, to reach the performance obtained against the SPM only. The eye diagrams of Figure \ref{fig:xpm_maps_sim}(c-f), obtained for $L = 150$ km, show that XPM acts as an additional noise source that manifests as random fluctuations in the transmitted bit, thus causing a closure of the diagram. The PNN still manages, at least partially, to counteract this phenomenon while being completely agnostic about the encoding of the pump signal. 

\section{Conclusion}
\label{sec:conclusion}

A Silicon Photonic Neural Network (PNN) is tested for linear and nonlinear effects equalization on 10 Gbaud PAM2 optical signals propagating in a multi-span link, with input power in fiber $P_{in}$ restoration at every span. In a linear propagation regime ($P_{in} = 0$ dBm), the PNN acts as a tunable photonic FIR filter, equalizing CD up to 200 km by tuning the phase weights only. On the other hand, in a nonlinear propagation regime ($P_{in} = 9$ dBm), SPM-induced distortions are compensated up to 450 km (deprived of CD contributions), this time with a primary role played by amplitude weights and the nonlinear activation function (here represented by the detection process). In both transmission regimes, the PNN maintains the BER below the $10^{-3}$ pre-FEC threshold with a very small power consumption ($290 \pm 40$ mW) produced solely by the thermal heaters. As a drawback, the PNN insertion loss remains within 18.4 dB and 22 dB (depending on the actual weights configuration). We foresee the implementation of integrated Semiconductor Optical Amplifiers (SOAs) to compensate for the losses and provide more versatility to the nonlinear stage when used in a saturated regime. Tests on these structures are ongoing.

\begin{figure}
    \centering
    \includegraphics[width=\linewidth]{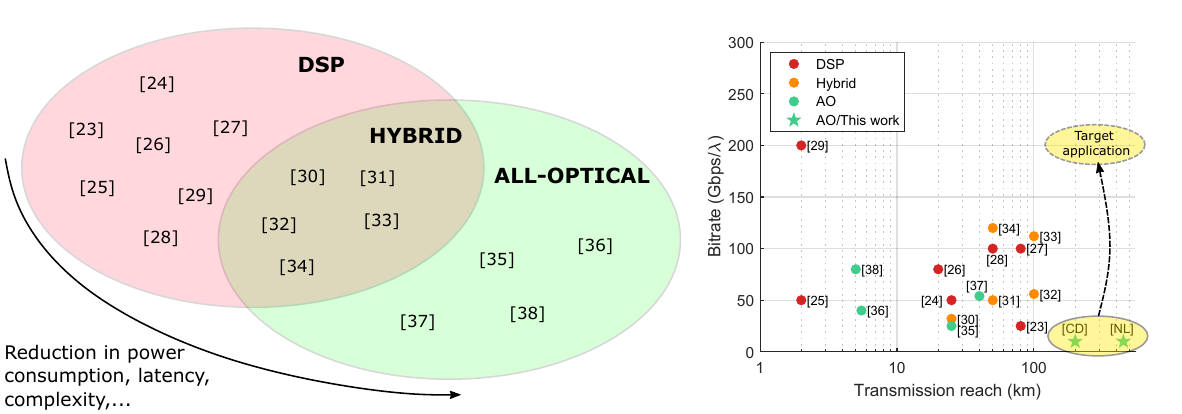}
    \caption{(left) Comparison of different approaches to signal equalization based on DSP (Digital Signal Processing), all-optical processing, and hybrid techniques. (right) Map of the best equalization performance in terms of bitrate and transmission reach obtained in the references (labels) via the different approaches. The results of the present work (green stars) are circled in yellow and projected towards the target application foreseen for the next-generation devices.}
    \label{fig:approach_comparison}
\end{figure}

Applying the presented feed-forward PNN to nonlinear equalization represents a true novelty, allowing a switch from a linear FIR interpretation to an all-optical time-delayed complex perceptron. This work is inserted in a broad research spectrum where different approaches, summarized in \textbf{Figure \ref{fig:approach_comparison}}, are used to tackle the problem of signal equalization. As of today, DSP-based technologies offer the highest reliability, robustness, and adaptability to different transmission scenarios \cite{cisco100G,xu2021joint,cisco200G, fang2024low, li2020low, wang2021100, cisco800G}, with the drawback of high power consumption, latency, and cost. A partial mitigation to this energy request and complexity is provided by hybrid approaches \cite{sackesyn2021experimental, wang2021signal, li2019optical, estebanez202256, sheng2024complex}, where the transmitted signal undergoes an optical processing, usually implemented as a Photonic Reservoir Computing, before being detected. The resulting sequences are then fed to a linear classifier for correct bit identification. A further step to DSP-free technologies is represented by all-optical approaches to signal equalization \cite{zuo2023integrated, nguyen2010tunable, brodnik2018extended, takiguchi2023integrated}, where this work is inserted. Minimized latency, complexity, and power consumption are the key-enabling features provided by our PNN, whose target application is a co-integration in an optical transceiver for telecom applications. In view of the next-generation devices' designs, an optimized PNN layout with 8-taps and a unitary delay of $\Delta t = 5 $ ps has been tested for nonlinear equalization in a 100 Gbaud scenario. In a simulated environment, the PNN provides SPM equalization, with up to 13 dB BER reduction for $P_{in} \sim 8$ dBm and $L \sim 350$ km. Beneficial effects are observed even against XPM in a scenario featuring two co-propagating signals, with up to an order of magnitude BER reduction for $L=250$ km.

\section{Experimental Section}
\subsection{Experimental setup and measurement process}
\label{sec:experimental_setup}
\begin{figure}
    \centering
    \includegraphics[width=1\linewidth]{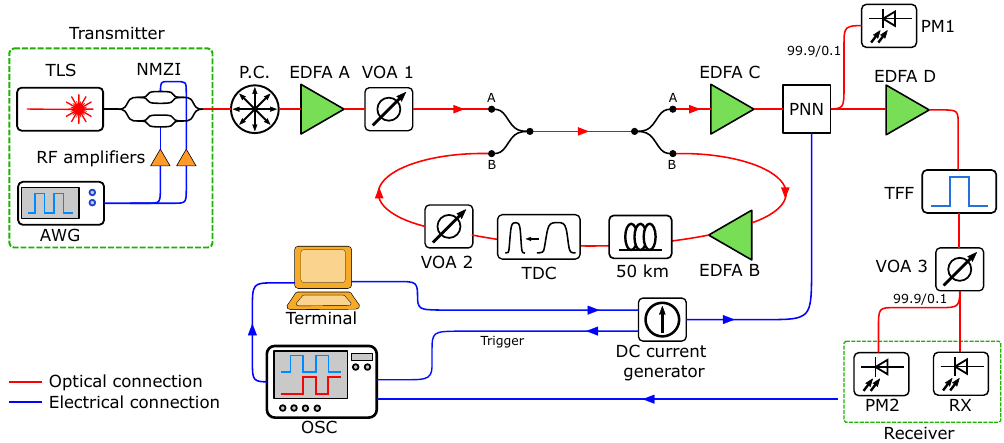}
    \caption{Experimental setup. TLS: Tunable Laser Source; NMZI: Nested Mach-Zehnder Interferometer; AWG: Arbitrary Waveform Generator; RF: Radio Frequency; P.C.: Polarization Controller; EDFA: Erbium-Doped Fiber Amplifier; VOA: Variable Optical Attenuator; TDC: Tunable Dispersion Compensator; PNN: Photonic Neural Network; TFF: Tunable Frequency Filter; RX: Fast Photodiode; PM: Power Monitor; OSC: Oscilloscope.}
    \label{fig:setup}
\end{figure}

\textbf{Figure \ref{fig:setup}} illustrates a schematic of the complete experimental setup. A continuous-wave (CW) tunable laser source (TLS) operating at 1550 nm is modulated to generate a 2-level Pulse Amplitude Modulated (PAM2) signal following the Non-Return-To-Zero (NRZ) paradigm. This signal is software-generated based on a periodic Pseudo-Random Binary Sequence (PRBS) of order 10, with a period of $2^{10}$ symbols. The sequence is fed to an Arbitrary Waveform Generator (AWG) with a 30 GHz bandwidth connected to two RF amplifiers with a 20 GHz bandwidth. These amplifiers drive a Nested Mach-Zehnder Interferometer (NMZI).

The modulated optical sequence passes through a polarization controller (PC), which is necessary due to the polarization sensitivity of the PNN device placed further in the line. An Erbium-Doped Fiber Amplifier (EDFA A) and a Variable Optical Attenuator (VOA 1) regulate the amount of optical power entering the propagation stage. This is implemented via two ultrafast optical switches (rise/fall time of 8 ns) used to fill and close the loop for propagation and, later, to empty it and direct light to the PNN. At every loop cycle, light propagates in a 50 km standard SMF-28 fiber span (corresponding to a propagation time of $\tau_F= 250$ $\mu$s), preceded by an EDFA (EDFA B) that restores the input power in the fiber up to 9 dBm. A Tunable Dispersion Compensator (TDC, max dispersion: -900 ps nm$^{-1}$) can be used to remove the accumulated CD, and a VOA 2 is used to fine-tune the optical gain in a single loop to 0 dB. The total propagation distance (i.e., the number of loop cycles) is set by properly delaying the change of state in the switches (discussed below). 

After the propagation, the distorted optical pulses reach an intermediate amplification stage (EDFA C) and proceed to the PNN for optical processing. Light enters and exits the chip via two tapered fibers in an edge coupling configuration, providing 4.5 dB of coupling losses. The total insertion loss depends on the selected weight configuration, with values ranging between 18.4 dB and 22 dB after the training procedures. A multi-channel DC generator sets the currents flowing through the micro-heaters on the chip, thus setting the amplitude and phase weights via the thermo-optic effect. The electrical power that generates a $\pi$ phase shift is measured to be, on average, $(24 \pm 2)$ mW. Again, the overall power consumption depends on the selected weight configuration, with average values of $(290 \pm 40)$ mW in the tested scenarios. At the output, the interference pattern is determined by the relative phase between the channels. Therefore, the micro-heater controlling the phase weight of the null-delay channel is left unconnected, choosing that channel as a reference. Instead, the corresponding DC generator line is used to trigger the oscilloscope. The PNN device is mounted on a Printed Circuit Board for electrical connections and placed in thermal contact with a temperature controller driven by a Peltier cell. During the measurements for unequalized transmission, the PNN is removed from the setup and replaced with a VOA (not depicted in Figure \ref{fig:setup}) set to 20 dB attenuation to mimic the device losses and maintain the overall OSNR budget.

After the PNN, a fused coupler with a $99.9/0.1$ splitting ratio sends a minor fraction of the optical power to a Power Monitor (PM1) to constantly check the device's alignment. The remaining fraction of optical power proceeds through  EDFA D, which compensates for the insertion losses, and subsequently, through a Tunable Frequency Filter (TFF) with a 30 GHz optical bandwidth. A final VOA (VOA 3) regulates the power (i.e., the Signal-To-Noise ratio) at the receiver stage, composed of a second Power Monitor (PM2) and a 20 GHz-bandwidth Photodiode (RX). The oscilloscope (OSC) featuring 16 GHz bandwidth for 80 GSa s$^{-1}$ sampling imposes the major bandwidth limitation in the setup.

A typical measurement procedure that leads to the separation loss function \cite{staffoli2025silicon} and BER evaluation occurs in different steps. Following the nomenclature appearing in Figure \ref{fig:setup}, at the beginning of each measurement, the switches are set in the $A \rightarrow A$ configuration, which is used for Back-To-Back (BTB) propagation. In this status, the loop is completely isolated from the BTB propagation line, and the EDFA B is left in an Amplified Spontaneous Emission (ASE) condition. Light from the transmitter is then diverted into the loop when the switches are set in configuration $A \rightarrow B$. EDFA B goes from an ASE condition to an active amplification of the incoming signal, which is associated with a transient in the device response \cite{meena2019mitigation}. Therefore, the $A \rightarrow B$ configuration is maintained for approximately $5 \tau_F$, during which light propagates through the dead-end loop and the EDFA's dynamics stabilize. Subsequently, the recirculation is enabled by switching to $B \rightarrow B$ configuration, which is held for a time entire multiple of $\tau_F$. After the propagation, the switches are set in $B \rightarrow A$ configuration, emptying the loop and directing light towards the PNN for further processing. The DC generator sends a triggering signal to the oscilloscope and a pre-set current array to the PNN device. The currents are held on for 220 $\mu$s to allow the local thermal relaxation of the micro-heaters and then immediately switched off to avoid the thermalization of the whole chip with consequent cross-talk between the different weights.

The measurement procedure relies on precise timing relations between triggering the switches and obtaining a controlled propagation, setting the PNN weights, and triggering the oscilloscope. The acquired trace is then processed via the terminal, evaluating the separation loss function and the BER according to the same modalities extensively described in \cite{staffoli2025silicon}.

Before the transmission tests, the PNN device is tested to characterize its operational parameters. In particular, we estimated the fixed amplitude weights $k_i$ for each channel, as shown in Equation \ref{eq:cperc}. The estimation begins with a condition of entirely constructive interference among all the channels. This condition was determined by setting $a_i=1$ for every $i$ and training the phase weights to maximize the output signal. Then, while maintaining fixed these phase weights, the channels were progressively closed. Namely, we set $a_1=0$, then $a_{1}=a_{2}=0$, and so on. We measured the output optical power after each closing. The experimental data is then provided to a PNN device model to fit the contributions of the individual channels. In the model, the actual circuit of the PNN (shown in Figure 1 of \cite{staffoli2025silicon}) was used. Table \ref{tab:ch_losses} reports the estimate values for $k_i$.

\begin{table}[h!]
    \centering
    \begin{tabular}{ccc|ccc}
        \hline
        Channel \#  &   Channel Loss  [dB]  &  $k_i$ &  Channel \#  &   Channel Loss [dB]  & $k_i$   \\
        \hline
        1   &   -22.5  & 0.0057  &   5   &  -15.1 & 0.0306  \\
        2   &   -18.7  & 0.0136 &   6   &  -16.7 & 0.0215 \\
        3   &   -17.4 & 0.0181 &   7   &  -17.4 & 0.0183 \\
        4   &   -16.5 & 0.0222  &   8   &  -17.8  & 0.0167 \\
        \hline
    \end{tabular}
    \vspace{2pt}
    \caption{Results of the PNN characterization procedure in terms of channel losses $k_i$ entering in Equation \eqref{eq:cperc}.}
    \label{tab:ch_losses}
\end{table}

\subsection{Simulation setup}
\label{sec:simulated_setup}
\begin{figure}
    \centering
    \includegraphics[width=\linewidth]{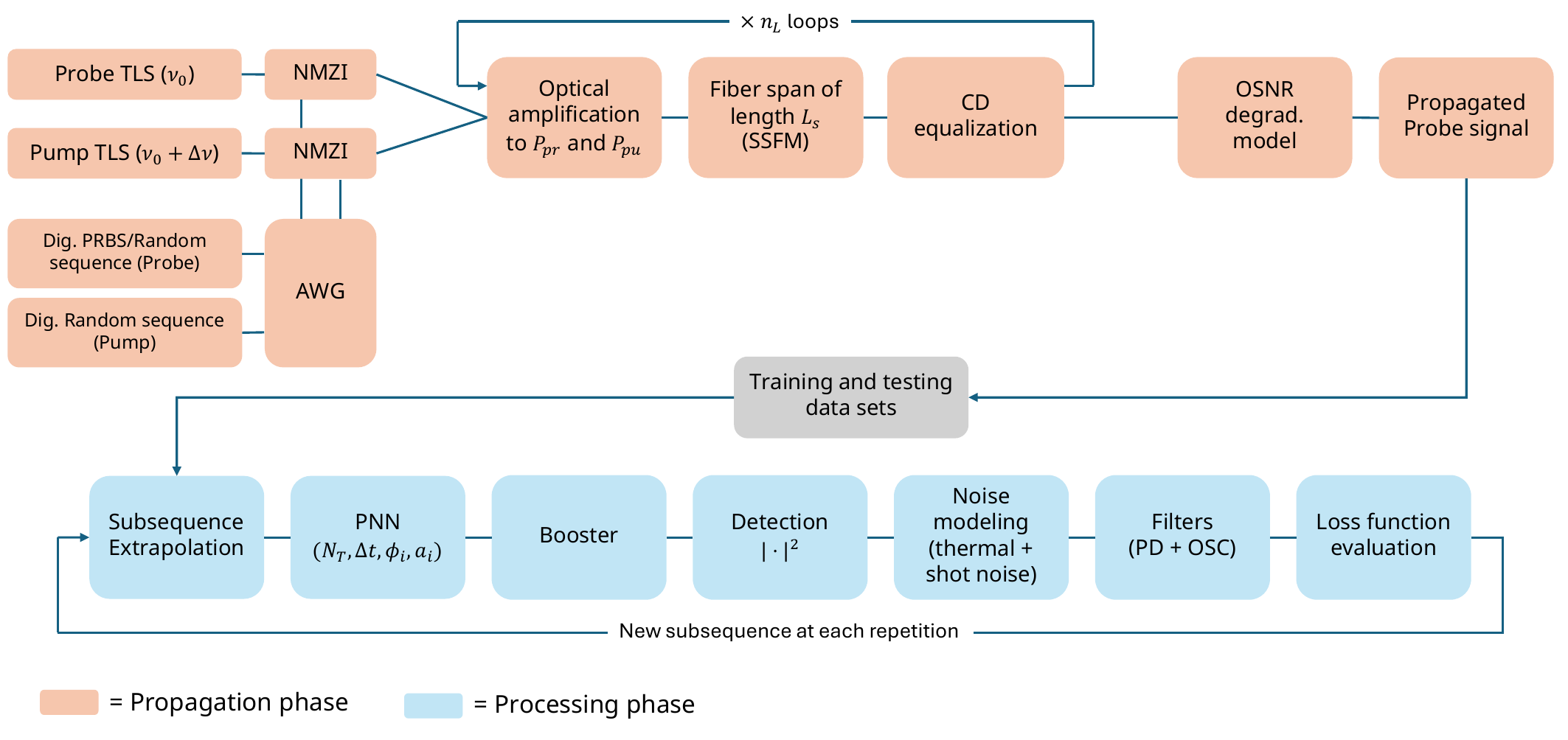}
    \caption{Block scheme for the simulated setup. A complete walk-through of the different blocks is provided in the text. TLS: Tunable Laser Source; AWG: Arbitrary Waveform Generator; NMZI: Nested Mach-Zehnder Modulator; SSMF: Split-Step Fourier Method; CD: Chromatic Dispersion; OSNR: Optical Signal-To-Noise Ratio; PNN: Photonic Neural Network; PD: Photodiode; OSC: Oscilloscope. }
    \label{fig:sim_setup_blocks}
\end{figure}

\textbf{Figure \ref{fig:sim_setup_blocks}} presents the block diagram of the setup reproduced via the simulation software. The proper parameters of each block and their used value or range of values are reported in \textbf{Table \ref{tab:params}}. The simulations happen in two phases: first occurs the propagation phase (orange blocks), where the optical signals are generated and propagated through the fiber loop. Secondly occurs the processing phase (blue blocks), where the propagated sequences are accessed, treated by the PNN, and detected, leading to the evaluation of the loss function. 

In the propagation phase, two Tunable Laser Sources (TLS) operating at the frequencies $\nu_0$ and $\nu_0 + \Delta \nu$ generate the Probe and Pump CW signals, respectively. These are then modulated via two Nested Mach-Zehnder Interferometers (NMZIs) driven by an Arbitrary Waveform Generator (AWG) with bandwidth $B_{AWG}$, resulting in a signal baud rate $B$. In our specific application, the probe is encoded with a 100 Gbps $2^{17}$ bits long sequence, which can be a PRBS of order 17, or a random sequence used to produce the training and testing data sets, respectively. On the other hand, the pump always encodes a random sequence of the same length and bitrate. A very high sampling frequency $f_{prop}$ is imposed on the two signals in this first phase to ensure accuracy during propagation. This is realized via a simulated fiber loop represented by a 3-block sequence, which is repeated for $n_L$ times to mimic a multi-span propagation. An optical amplification stage implemented as a rescaling factor within the loop restores the optical power to $P_{pr}$ and $P_{pu}$ at every recirculation. The propagation through a fiber span of length $L_{s}$ is reproduced via the Split-Step Fourier Method (SSFM) \cite{agrawal2013nonlinearSSFM}. The nomenclature for the SSFM parameters proposed in the text and Table \ref{tab:params} follows the one presented in \cite{agrawal2013nonlinearCh1}. The subsequent CD equalization stage is described by the operator $\exp \{ -j L_{s} \beta_2 \omega^2/2  - j L_{s} \beta_3 \omega^3/6\} $ in the frequency domain, representing the inverse of the linear propagation term in a lossless fiber. At this point, the pump signal is discarded since it only co-propagates with the probe and triggers the XPM effect. On the other hand, the probe signal is added with optical noise according to the model proposed in \cite{agrawal2021fiber,chomycz2009osnr} to account for the OSNR degradation imposed by the optical amplification at each loop. In particular, after a $n_l$ loop propagation, the probe signal's spectrum is added with Gaussian noise to match the OSNR condition predicted by
\begin{equation}
    OSNR = P_{out} - l_s - NF - 10\log_{10}(n_l) - 10\log_{10}( h\nu_0 \Delta\nu_{bw}).
\end{equation}
Here, $P_{out}$ is the output power from the amplification stage (i.e., $P_{pr}$), $l_s$ is the loss accumulated in a loop as the sum of the contributions from fiber, TDC, VOA 3, and switches, NF is the noise figure of the amplifiers, $\Delta \nu_{bw}$ is the bandwidth chosen for OSNR evaluation, and $h$ is the plank constant. The so-obtained propagated probe signal is stored for future utilization, and this step ends the propagation phase. The process described so far is repeated for different transmission scenarios in terms of $P_{pr}$, $P_{pu}$, and propagation distance $L = L_s \times n_L$, thus creating the training and testing data sets.

Subsequently, the processing phase starts, consisting of a whole run through the steps described in the blue rectangles of Figure \ref{fig:sim_setup_blocks}. Each run leads to a loss function evaluation, and thus, it is iteratively repeated in the context of PNN training or testing. Each run starts by accessing a stored propagated sequence and extrapolating a sub-sequence according to the procedures described in Figure \ref{fig:datasets_explanation}. Then, occurs the processing of a parametric PNN featuring $N_{T}$ taps and $\Delta t$ unitary delay, and equipped with amplitude and phase weights. Spiral losses $\alpha_s$ are accounted too, and the relative delay between the $N_{T}$ copies of the input signal is manually imposed by time-shifting them. The resulting sequence is then applied with a rescaling factor to boost the optical power to the desired value $P_{RX}$ (during the testing phase) or to emulate a constant-gain amplifier (during the training phase). After the detection, implemented with the square-modulus operation and a conversion gain $R_{PD}$ (Watts-to-Volts), the resulting signal $\vert s \vert^2$ (expressed in mV is added with Gaussian noise according to the model suggested in \cite{agrawall2010receivers}. Its variance is given by a thermal $\sigma_T$ and a shot noise $\sigma_{s}$ contribution, namely
\begin{equation}
    \sigma^2 = \sigma_{S}^2 \vert s \vert^2 + \sigma_T^2
\end{equation}
The code calibration with 10 Gbps signals provides the model parameters' value. The noise-added signal undergoes a double filtering process implemented with a 5th order Bessel Filter accounting for the photodiode bandwidth $B_{PD}$ and the oscilloscope bandwidth $B_{OSC}$. A down-sampling at a frequency $f_{osc}$ is then operated, together with a $n_{vert}$-bit vertical quantization of the full scale $V_{max}$. The 2-sample Separation Loss function and BER are finally evaluated on the so-obtained output sequence.

\renewcommand{\arraystretch}{1.5}
\begin{table}[t]
    \centering
    \begin{tabular}{lcc}
    \hline
    Parameter Description & Symbol & Value \\
    \hline
    Baud rate & $B$ & 10 to 100 Gbps \\
    Probe Laser Frequency & $\nu_{0}$ & $193.55$ THz \\
    Detuning between probe and pump laser sources & $ \Delta \nu$ & 50 GHz \\
    Sampling frequency for propagation phase & $f_{prop}$ & $B \times 80$ \\
    Arbitrary Waveform Generator bandwidth & $B_{AWG}$ & $3 \times B$ \\
    Losses in a single loop round-trip & $l_s$ & 18 dB \\
    Noise figure in OSNR degradation model & $NF$ & 5 dB \\
    Bandwidth for OSNR evaluation & $\Delta \nu_{bw}$ & $1.2\times B$ \\
    Photodiode bandwidth & $B_{PD}$ & $2 \times B$ \\
    Photodiode Conversion Gain & $R_{PD}$ & 24 V W$^{-1}$ \\
    Oscilloscope sampling frequency & $f_{osc}$ & $B \times 8$ \\
    Oscilloscope vertical full scale & $V_{max}$ & 160 mV \\
    Oscilloscope vertical resolution & $n_{vert}$ & 8 bits \\
    Oscilloscope bandwidth & $B_{osc}$ & $1.6 \times B$ \\
    Number of loops & $n_L$ & 0 to 9 \\
    Fiber span length & $L_s$ & 50 km \\
    Fiber dispersion & $D$ & 18 ps nm$^{-1}$ km$^{-1}$ \\
    Group velocity dispersion coefficient & $\beta_2$ & 0.022 ps$^2$ m$^{-1}$ \\
    Third-order dispersion coefficient & $\beta_3$ & 0.1 ps$^3$ km$^{-1}$ \\
    Effective mode area & $A_{eff}$ & $80 \times 10^{-12}$ m$^2$ \\
    Kerr coefficient & $n_{2}$ & $2.7 \times 10^{-20}$ m$^{2}$ W$^{-1}$ \\
    Fiber Attenuation constant & $\alpha_{\text{dB}}$ & 0.2 dB km$^{-1}$ \\
    Split-Step Fourier Method propagation step & $h_s$ & 138.8 m \\
    Thermal noise contribution & $\sigma_{T}^2$ & 0.65 mV$^2$\\
    Shot noise contribution & $\sigma_{S}^2$ & 0.05 mV \\
    Delay lines losses per unit length & $\alpha_s$ & 2 dB cm$^{-1}$ \\ 
    \hline
    \end{tabular}
    \caption{Parameters used in the simulated setup. Section \ref{sec:simulated_setup} provides a more detailed description of each table entry.}
    \label{tab:params}
\end{table}

\medskip
\textbf{Supporting Information} \par 
Supporting Information is available from the Wiley Online Library or from the author.

\medskip
\textbf{Acknowledgements} \par 
We acknowledge Fondazione CARITRO, which supported this work via a POC R2M 2024 project. We acknowledge Dr. Alessio Lugnan, Dr. Stefano Biasi, and Prof. Alberto Gatto (PoliMi) for the fruitful discussions.

\medskip

%
\bibliographystyle{MSP}
\bibliography{bibliography}

\begin{thebibliography}{10}
\providecommand{\url}[1]{\texttt{#1}}
\providecommand{\urlprefix}{URL }

\bibitem{zhong2018digital}
K.~Zhong, X.~Zhou, J.~Huo, C.~Yu, C.~Lu, A.~P.~T. Lau,
\newblock Digital {S}ignal {P}rocessing for {S}hort-{R}each {O}ptical {C}ommunications: {A} {R}eview of {C}urrent {T}echnologies and {F}uture trends,
\newblock \emph{Journal of Lightwave Technology} \textbf{2018}, \emph{36}, 2 377.

\bibitem{guifang2009recent}
G.~Li,
\newblock Recent advances in coherent optical communication,
\newblock \emph{Adv. Opt. Photon.} \textbf{2009}, \emph{1}, 2 279.

\bibitem{liu2014digital}
X.~Liu, S.~Chandrasekhar, P.~J. Winzer,
\newblock Digital signal processing techniques enabling multi-{T}b/s superchannel transmission: an overview of recent advances in {DSP}-enabled superchannels,
\newblock \emph{IEEE Signal Processing Magazine} \textbf{2014}, \emph{31}, 2 16.

\bibitem{zhou2014advanced}
X.~Zhou, L.~Nelson,
\newblock Advanced {DSP} for 400 {G}b/s and beyond optical networks,
\newblock \emph{Journal of lightwave technology} \textbf{2014}, \emph{32}, 16 2716.

\bibitem{sonntag2006digital}
J.~L. Sonntag, J.~Stonick,
\newblock A digital clock and data recovery architecture for multi-gigabit/s binary links,
\newblock \emph{IEEE Journal of Solid-State Circuits} \textbf{2006}, \emph{41}, 8 1867.

\bibitem{chang2010forward}
F.~Chang, K.~Onohara, T.~Mizuochi,
\newblock Forward error correction for 100 {G} transport networks,
\newblock \emph{IEEE Communications Magazine} \textbf{2010}, \emph{48}, 3 S48.

\bibitem{huang2022performance}
L.~Huang, Y.~Xu, W.~Jiang, L.~Xue, W.~Hu, L.~Yi,
\newblock Performance and {C}omplexity {A}nalysis of {C}onventional and {D}eep {L}earning {E}qualizers for the {H}igh-{S}peed {IMDD} {PON},
\newblock \emph{Journal of Lightwave Technology} \textbf{2022}, \emph{40}, 14 4528.

\bibitem{agrawal2013nonlinearCh1}
G.~Agrawal,
\newblock Chapter 1 - {I}ntroduction,
\newblock In \emph{Nonlinear {F}iber {O}ptics ({F}ifth {E}dition)}, Optics and Photonics, 1--25. Academic Press, Boston, fifth edition edition, \textbf{2013}.

\bibitem{frey2017estimation}
F.~Frey, R.~Elschner, J.~K. Fischer,
\newblock Estimation of trends for coherent {DSP} {ASIC} power dissipation for different bitrates and transmission reaches,
\newblock In \emph{Photonic Networks; 18. ITG-Symposium}. VDE, \textbf{2017} 1--8.

\bibitem{cheng2019comparison}
J.~Cheng, C.~Xie, Y.~Chen, X.~Chen, M.~Tang, S.~Fu,
\newblock Comparison of {C}oherent and {IMDD} {T}ransceivers for {I}ntra {D}atacenter {O}ptical {I}nterconnects,
\newblock In \emph{2019 Optical Fiber Communications Conference and Exhibition (OFC)}. \textbf{2019} 1--3.

\bibitem{staffoli2025silicon}
E.~Staffoli, G.~Maddinelli, L.~Pavesi,
\newblock A {S}ilicon {P}hotonic {N}eural {N}etwork for {C}hromatic {D}ispersion {C}ompensation in 20 {G}bps {PAM4} {S}ignal at 125 km and its {S}calability up to 100 {G}bps,
\newblock \emph{J. Lightwave Technol.} \textbf{2025}, \emph{43}, 2 557.

\bibitem{staffoli2023equalization}
E.~Staffoli, M.~Mancinelli, P.~Bettotti, L.~Pavesi,
\newblock Equalization of a 10 {G}bps {IMDD} signal by a small silicon photonics time delayed neural network,
\newblock \emph{Photon. Res.} \textbf{2023}, \emph{11}, 5 878.

\bibitem{staffoli2024chromaticPW}
E.~Staffoli, G.~Maddinelli, M.~Mancinelli, P.~Bettotti, L.~Pavesi,
\newblock {Chromatic dispersion compensation via an all-optical perceptron},
\newblock In G.~Li, K.~Nakajima, A.~K. Srivastava, editors, \emph{Next-Generation Optical Communication: Components, Sub-Systems, and Systems XIII}, volume 12894. International Society for Optics and Photonics, SPIE, \textbf{2024} 128940G,
\newblock \urlprefix\url{https://doi.org/10.1117/12.2692979}.

\bibitem{marciano2025chromatic}
P.~R.~N. Marciano, E.~Staffoli, G.~Maddinelli, M.~S. Coelho, L.~C.~B. Silva, J.~A.~L. Silva, M.~J. Pontes, M.~E.~V. Segatto, L.~Pavesi,
\newblock Chromatic {D}istortion {P}recompensation in {OFDM}-{B}ased {O}ptical {S}ystems {T}hrough an {I}ntegrated {S}ilicon {P}hotonic {N}eural {N}etwork,
\newblock \emph{Journal of Lightwave Technology} \textbf{2025}, \emph{43}, 7 3034.

\bibitem{bishop2006pattern}
C.~M. Bishop, N.~M. Nasrabadi,
\newblock \emph{Pattern recognition and machine learning}, volume~4, chapter~4,
\newblock Springer, \textbf{2006}.

\bibitem{mancinelli2022photonic}
M.~Mancinelli, D.~Bazzanella, P.~Bettotti, L.~Pavesi,
\newblock A photonic complex perceptron for ultrafast data processing,
\newblock \emph{Scientific Reports} \textbf{2022}, \emph{12}, 1 1.

\bibitem{hui1999cross}
R.~Hui, K.~R. Demarest, C.~T. Allen,
\newblock Cross-phase modulation in multispan {WDM} optical fiber systems,
\newblock \emph{Journal of lightwave Technology} \textbf{1999}, \emph{17}, 6 1018.

\bibitem{agrawal2013nonlinearSSFM}
G.~Agrawal,
\newblock Chapter 2 - {P}ulse {P}ropagation in {F}ibers,
\newblock In \emph{Nonlinear Fiber Optics (Fifth Edition)}, Optics and Photonics, 27--56. Academic Press, Boston, fifth edition edition, \textbf{2013}.

\bibitem{agrawal2021fiber}
G.~P. Agrawal,
\newblock Chapter 7 - {L}oss {M}anagement,
\newblock In \emph{Fiber‐Optic Communication Systems}, 235--275. John Wiley \& Sons, Ltd,
\newblock ISBN 9781119737391, \textbf{2021}.

\bibitem{chomycz2009osnr}
B.~Chomycz,
\newblock Optical {S}ignal to {N}oise {R}atio,
\newblock In \emph{Planning Fiber Optics Networks}, chapter~3. McGraw-Hill, New York, 1st edition, \textbf{2009}.

\bibitem{efron1992bootstrap}
B.~Efron,
\newblock Bootstrap methods: another look at the jackknife,
\newblock In \emph{Breakthroughs in statistics: Methodology and distribution}, 569--593. Springer, \textbf{1992}.

\bibitem{ader2008advising}
H.~Ad{\`e}r, D.~Hand, G.~Mellenbergh,
\newblock \emph{Advising on {R}esearch {M}ethods: {A} {C}onsultant's {C}ompanion},
\newblock Johannes Van Kessel Publishing, \textbf{2008}.

\bibitem{cisco100G}
100{G}{B}{A}{S}{E} {Q}{S}{F}{P}-100{G} {M}odules {D}ata {S}heet --- cisco.com,
\newblock \url{https://www.cisco.com/c/en/us/products/collateral/interfaces-modules/transceiver-modules/datasheet-c78-736282.html},
\newblock [Accessed 21-03-2025].

\bibitem{xu2021joint}
Z.~Xu, C.~Sun, J.~H. Manton, W.~Shieh,
\newblock Joint equalization of linear and nonlinear impairments for pam4 short-reach direct detection systems,
\newblock \emph{IEEE Photonics Technology Letters} \textbf{2021}, \emph{33}, 9 425.

\bibitem{cisco200G}
{C}isco {T}ransceiver {M}odules - {C}isco 200{G} {Q}{S}{F}{P}56 {C}ables and {T}ransceiver {M}odules {D}ata {S}heet --- cisco.com,
\newblock \url{https://www.cisco.com/c/en/us/products/collateral/interfaces-modules/transceiver-modules/nb-06-200g-qsfp56-cables-trans-mod-ds-cte-en.html},
\newblock [Accessed 21-03-2025].

\bibitem{fang2024low}
X.~Fang, M.~Bi, Z.~Li, L.~Jin, G.~Yang, J.~Shang, M.~Hu,
\newblock Low complexity deep neural network equalizer based on the multi-source domain transfer learning in {IMDD} system,
\newblock \emph{Optics Express} \textbf{2024}, \emph{32}, 19 33004.

\bibitem{li2020low}
D.~Li, H.~Song, W.~Cheng, M.~Cheng, S.~Fu, M.~Tang, D.~Liu, L.~Deng,
\newblock Low-complexity equalization scheme for suppressing {FFE}-enhanced in-band noise and {ISI} in 100 {G}bps {PAM4} optical {IMDD} system,
\newblock \emph{Optics letters} \textbf{2020}, \emph{45}, 9 2555.

\bibitem{wang2021100}
H.~Wang, P.~Torres-Ferrera, G.~Rizzelli, V.~Ferrero, R.~Gaudino,
\newblock 100 {G}bps/$\lambda$ {C}-band {CD} digital pre-compensated and direct-detection links with simple non-linear compensation,
\newblock \emph{IEEE Photonics Journal} \textbf{2021}, \emph{13}, 4 1.

\bibitem{cisco800G}
{C}isco {T}ransceiver {M}odules - {C}isco {Q}{S}{F}{P}-{D}{D}800 {T}ransceiver {M}odules {D}ata {S}heet --- cisco.com,
\newblock \url{https://www.cisco.com/c/en/us/products/collateral/interfaces-modules/transceiver-modules/qsfp-dd800-transceiver-modules-ds.html},
\newblock [Accessed 21-03-2025].

\bibitem{sackesyn2021experimental}
S.~Sackesyn, C.~Ma, J.~Dambre, P.~Bienstman,
\newblock Experimental realization of integrated photonic reservoir computing for nonlinear fiber distortion compensation,
\newblock \emph{Optics Express} \textbf{2021}, \emph{29}, 20 30991.

\bibitem{wang2021signal}
S.~Wang, N.~Fang, L.~Wang,
\newblock Signal recovery based on optoelectronic reservoir computing for high speed optical fiber communication system,
\newblock \emph{Optics Communications} \textbf{2021}, \emph{495} 127082.

\bibitem{li2019optical}
S.~Li, S.~Pachnicke,
\newblock Optical equalization using photonic reservoir computing with optical analog signal injection,
\newblock In \emph{Asia Communications and Photonics Conference}. Optica Publishing Group, \textbf{2019} T4G--5.

\bibitem{estebanez202256}
I.~Est{\'e}banez, S.~Li, J.~Schwind, I.~Fischer, S.~Pachnicke, A.~Argyris,
\newblock {56} {GB}aud {PAM-4} 100 km transmission system with photonic processing schemes,
\newblock \emph{Journal of Lightwave Technology} \textbf{2022}, \emph{40}, 1 55.

\bibitem{sheng2024complex}
W.~Sheng, C.~Liu, J.~Xiao, L.~Sun, Y.~Cai, H.~Fu, Q.~Li, G.~Ning~Liu,
\newblock Complex-valued recurrent neural network equalizer with low complexity for a 120-{G}bps 50-km optical {PAM-4} {IM/DD} system,
\newblock \emph{Optics Express} \textbf{2024}, \emph{32}, 16 27624.

\bibitem{zuo2023integrated}
X.~Zuo, L.~Pei, B.~Bai, J.~Wang, J.~Zheng, T.~Ning, F.~Dong, Z.~Zhao,
\newblock Integrated silicon photonic reservoir computing with {PSO} training algorithm for fiber communication channel equalization,
\newblock \emph{Journal of Lightwave Technology} \textbf{2023}, \emph{41}, 18 5841.

\bibitem{nguyen2010tunable}
H.~M. Nguyen, K.~Igarashi, K.~Katoh, K.~Kikuchi,
\newblock Tunable optical equalizer for 40-{G}bps intensity-modulated signal using {PLC}-based finite-impulse-response filter,
\newblock In \emph{36th European Conference and Exhibition on Optical Communication}. IEEE, \textbf{2010} 1--3.

\bibitem{brodnik2018extended}
G.~M. Brodnik, C.~Pinho, F.~Chang, D.~J. Blumenthal,
\newblock Extended reach 40km transmission of {C}-band real-time 53.125 {G}bps {PAM-4} enabled with a photonic integrated tunable lattice filter dispersion compensator,
\newblock In \emph{2018 Optical Fiber Communications Conference and Exposition (OFC)}. IEEE, \textbf{2018} 1--3.

\bibitem{takiguchi2023integrated}
K.~Takiguchi,
\newblock Integrated-optic chromatic dispersion compensator with completely passive operation and wide operational bandwidth,
\newblock \emph{Optics Continuum} \textbf{2023}, \emph{2}, 12 2529.

\bibitem{meena2019mitigation}
D.~Meena, K.~Sarath, F.~Francis, E.~Dipin, T.~Srinivas,
\newblock Mitigation of {EDFA} transient effects in variable duty cycle pulsed signals,
\newblock \emph{Defence Technology} \textbf{2019}, \emph{15}, 3 276.

\bibitem{agrawall2010receivers}
G.~P. Agrawal,
\newblock Chapter 4 - {O}ptical {R}eceivers,
\newblock In \emph{Fiber‐Optic Communication Systems}, 128--181. John Wiley \& Sons, Ltd,
\newblock ISBN 9780470918524, \textbf{2010}.

\end{thebibliography}






\end{document}